\documentclass[aps,floats]{revtex4}
\usepackage{tabularx,graphicx}
\usepackage{epsfig}
\usepackage{amsmath}
\usepackage{bbm}
\usepackage{graphicx}

\begin{document}

\title{Propagating, evanescent, and localized states in carbon nanotube-graphene
junctions}
\author{J. Gonz\'alez$^a$, F. Guinea$^b$ and J. Herrero$^a$  \\}
\address{$^a$Instituto de Estructura de la Materia,
        Consejo Superior de Investigaciones Cient\'{\i}ficas, Serrano 123,
        28006 Madrid, Spain\\
        $^b$Instituto de Ciencia de Materiales de Madrid,
        Consejo Superior de Investigaciones Cient\'{\i}ficas, Cantoblanco,
        Madrid, Spain}

\date{\today}

\begin{abstract}

We study the electronic structure of the junctions between a single graphene 
layer and carbon nanotubes, using a tight-binding model and the continuum 
theory based on Dirac fermion fields. The latter provides a unified description 
of different lattice structures with curvature, which is always localized at 
six heptagonal carbon rings around each junction. When these are evenly 
spaced, we find that it is possible to curve the planar lattice into armchair 
$(6n,6n)$ as well as zig-zag $(6n,0)$ nanotubes. We show that the junctions 
fall into two different classes, regarding the low-energy electronic behavior. 
One of them, constituted by the junctions made of the armchair nanotubes
and the zig-zag $(6n,0)$ geometries when $n$ is a multiple of 3, is characterized 
by the presence of two quasi-bound states at the Fermi level, which are absent 
for the rest of the zig-zag nanotubes. These states, localized at the junction, 
are shown to arise from the effective gauge flux induced by the 
heptagonal carbon rings, which has a direct reflection in the local density 
of states around the junction. Furthermore, we also analyze the
band structure of the arrays of junctions, finding out that they can also be
classified into two different groups according to the low-energy 
behavior. In this regard, the arrays made of armchair and $(6n,0)$ nanotubes
with $n$ equal to a multiple of 3 are characterized by the presence of a 
series of flat bands, whose number grows with the length of the nanotubes. 
We show that such flat bands have their origin in the formation of states
confined to the nanotubes, with little overlap in the region between the 
junctions. This is explained in the continuum theory from the possibility
of forming standing waves in the mentioned nanotube geometries, as a
superposition of modes with opposite momenta and the same quantum numbers 
under the $C_{6v}$ symmetry of the junction.

\end{abstract}

\maketitle

\section{Introduction}
The recent isolation of graphene layers a single atom
thick\cite{Netal04,Netal05,NGPNG08} has lead to a great deal of
activity, because of their novel electronic properties and potential
applications. The lattice structure of graphene is determined by the
$sp^2$ coordination between neighboring carbon atoms. Each carbon
atom has three nearest neighbors, leading to planar honeycomb
lattice. With small modifications, the same structure describes
other carbon allotropes, such as the fullerenes and the carbon nanotubes.

In this paper we study the simplest systems which combine two of these
allotropes: the junctions between a single graphene layer and carbon
nanotubes. A regular array of closely spaced armchair nanotubes attached 
to a graphene layer has already been studied\cite{jap}, and related systems 
are being considered for their potential applications\cite{F08}. We undertake
here the investigation of junctions made of nanotubes with different 
chiralities, which have in common a transition from the planar to the tubular 
geometry mediated by the presence of six heptagonal carbon rings. These induce 
the negative curvature needed to bend the honeycomb carbon lattice at the 
junction, playing a kind of dual role to that of the pentagonal carbon
rings in the fullerene cages\cite{ggv}.

We analyze first the electronic properties of a single junction between
a carbon nanotube and a graphene layer. We discuss the possible structures 
of this type, concentrating on geometries where the heptagonal rings are 
evenly spaced around the junction. The nanotubes can be then either armchair
$(6n,6n)$ 
or zig-zag with $(6n, 0)$ geometry (that is, with $6n$ hexagonal rings around 
the tube). We calculate their electronic structure, using the tight-binding 
model based on the $\pi$ orbitals of the carbon atoms widely applied to carbon
allotropes with $sp^2$ coordination. 

Paying attention to the local density of 
states, we find that the junctions fall into two different classes, depending 
on the behavior in the low-energy regime. One of the classes, comprising the 
junctions made of armchair and $(6n, 0)$ nanotubes when $n$ is a multiple of 3, 
is characterized by the presence of a peak in the density of states close to 
the Fermi level. The peak is absent in the other class, formed by the junctions
made with the rest of zig-zag geometries. In general, the density of states 
tends to be depleted in the junction at low energies, with peaks above and 
below the Fermi level marking the threshold for the propagation of new states
across the junction.     

We present next a continuum description, based on the formulation of Dirac
fermion fields in the curved geometry, which allows us to characterize the 
general properties of the junction, and which is consistent with the previous 
discrete analysis. Thus, we see that the peak at the Fermi level in the local 
density of states
is in general a reflection of the existence of quasi-bound states (zero modes)
for the Dirac equation in the curved space of the junction. It is known that
the topological defects of the honeycomb lattice (pentagonal and heptagonal
rings) induce an effective gauge field in the space of the two Dirac points 
of the planar graphene lattice\cite{ggv2}. It turns out that the effective 
magnetic flux is enough to localize two states at the junctions made of 
armchair or $(6n, 0)$ nanotubes when $n$ is a multiple of 3. At low energies, 
however, the generic behavior is given by evanescent states, which arise from 
the matching of modes with nonvanishing angular momentum and have exponential 
decay in the nanotube.

We finally apply our computational framework to the analysis of the band
structure of the arrays of nanotube-graphene junctions. Considering the behavior
of the low-energy bands close to the Fermi level, we find that the arrays also
fall into two different classes. The arrays made of armchair nanotubes or 
$(6n, 0)$ nanotubes with $n$ equal to a multiple of 3 tend to have a series of
flat bands close to the Fermi level, while the arrays made with the rest of 
zig-zag nanotubes have all the bands dispersing at low energies. Such a 
different behavior has its origin in the existence of states confined in
the nanotube side of the junction. We find that this feature can also be
explained in the context of the continuum model. The armchair and the $(6n, 0)$ 
geometries with $n$ equal to a multiple of 3 allow for the formation of 
standing waves between the junction and the other end of the tube. This is the 
mechanism responsible for the confinement of the states in the nanotubes and 
the consequent development of the flat bands, whose number grows at low 
energies with the length of the nanotube, in agreement with the predictions of
the continuum theory.

\section{Tight-binding approach to carbon nanotube-graphene structures}

\subsection{Lattice structure}
\begin{figure}[h]

\vspace{0.5cm}

\begin{center}
\mbox{\epsfysize 3cm \epsfbox{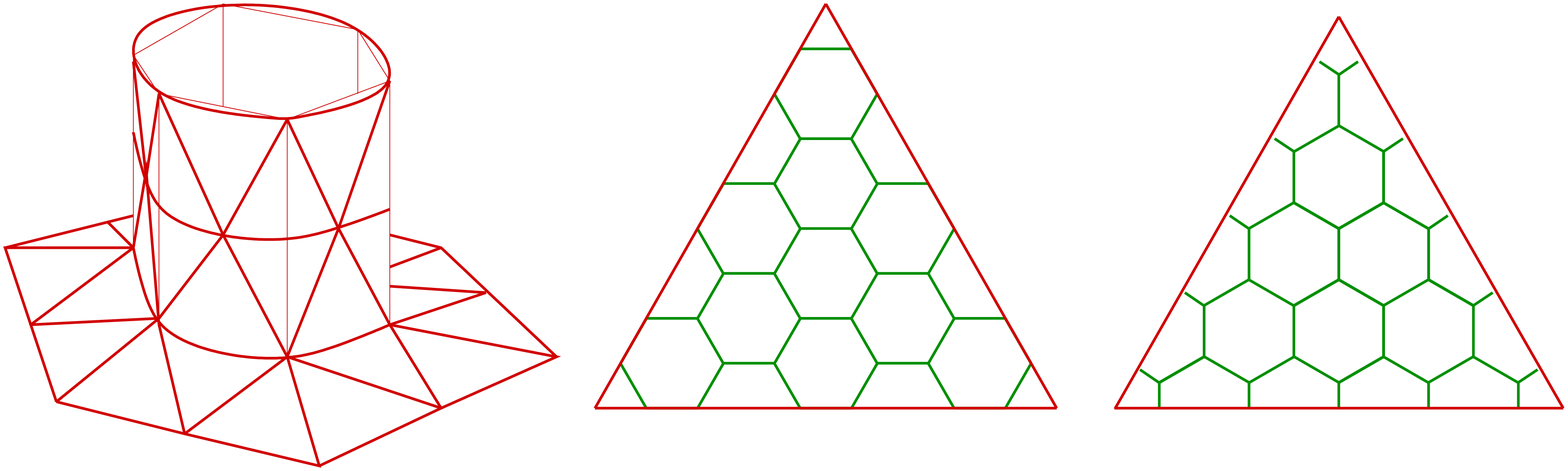}}   \\
 \hspace{0.1cm}  (a) \hspace{3.0cm} (b)  \hspace{2.7cm}  (c)

\end{center}
\caption{(a) Sketch of carbon nanotube attached to a graphene plane. The
building blocks of the structure are triangles which include many carbon atoms.
The orientation of the bonds of the honeycomb lattice in the basic triangle
may give rise to armchair (b) or zig-zag nanotubes (c). The threefold
coordination of the carbon atoms induces the existence of six heptagonal rings
at the ends of the sides of the hexagonal prism contacting the plane in (a).}
\label{one}
\end{figure}

Our first aim is to analyze the density of states of a semi-infinite
nanotube attached to a graphene layer in the tight-binding
approximation. The possible setups that we will consider, keeping
the threefold coordination of the carbon atoms, are sketched in Fig.
\ref{one}. The structures can be wrapped by the graphene hexagonal
lattice, with the exception of the six points where the sides of the
hexagonal prism (which describes the nanotube) intersect the plane.
The threefold coordination of the carbon atoms requires the
existence of sevenfold rings at those positions.

We describe the electronic states in the structures shown in Fig. \ref{one}
by means of a nearest-neighbor tight-binding model. In general the relaxation
of elastic energy will modify the bond lengths at the junction, depending on
the nanotube radius. We will assume that this relaxation does not change
significantly the electronic behavior. In this respect, a tight-binding model
based on the $\pi $ carbon orbitals is well-suited for the purpose of
discerning the extended or localized character of the different electronic
states. Our main achievement will be to assign the different features in the
local density of states to the behavior of the electronic states near the 
nanotube-graphene junctions. To this aim, we have actually checked that slight 
modulations of the transfer integral $t$ near the junctions do not produce 
significant changes in the results shown in what follows.

\subsection{Electronic densities of states}

We concentrate on the analysis of geometries where the six heptagonal carbon
rings are evenly spaced around the junction as in Fig. \ref{one}. This 
constrains the possible chiralities of the nanotubes, that can be then either
armchair $(6n,6n)$ or zig-zag $(6n,0)$, with the number $n$ running over all
the integers. Nanotubes in which the carbon sheet is wrapped with helicity
can be also attached at the expense of introducing an irregular distribution
of the heptagonal rings. Anyhow, we expect that the rules explaining the 
different features in the density of states are universal enough to hold even 
in these more general cases.

We have obtained the spectra of different types of hybrid structures by
diagonalization of the tight-binding hamiltonian for very large lattices,
with up to $\approx 50,000$ carbon atoms in the graphene part and
$\approx 40,000$ in the nanotube side. Given that the whole geometry has
$C_{6v}$ symmetry, we have classified the energy eigenstates into six groups
according to the eigenvalue $q$ under a rotation of $\pi /3$.
The nature of each electronic state is given in general by its behavior
at the nanotube-graphene junction. For this reason, we have characterized
the hybrid structures in terms of the local density of states averaged over 
a circular ring of atoms at the end of the nanotube close to the junction.

\begin{figure}
\begin{center}
\mbox{
\epsfxsize 4cm \epsfbox{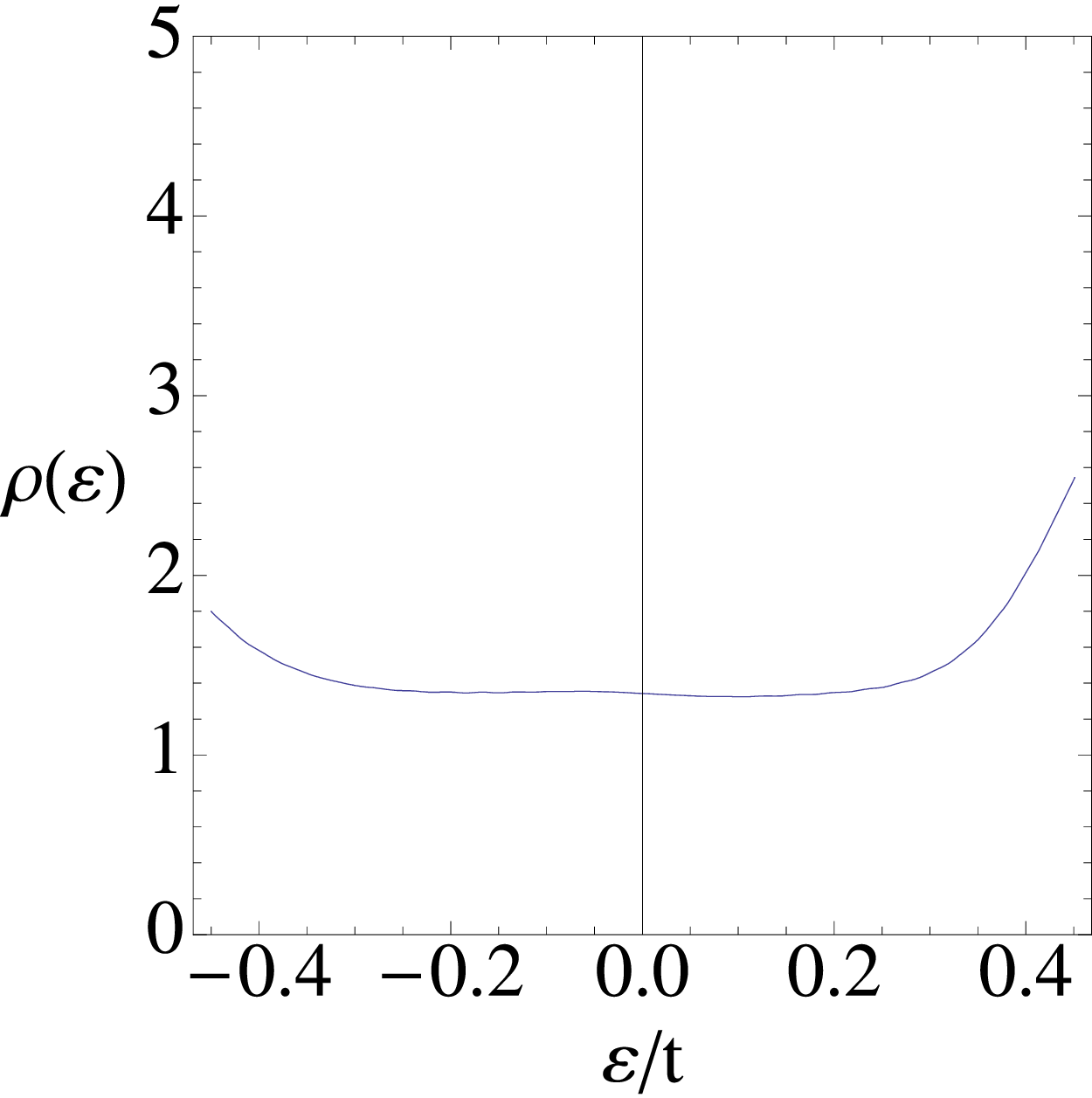}  \hspace{0.5cm}
\epsfxsize 4cm   \epsfbox{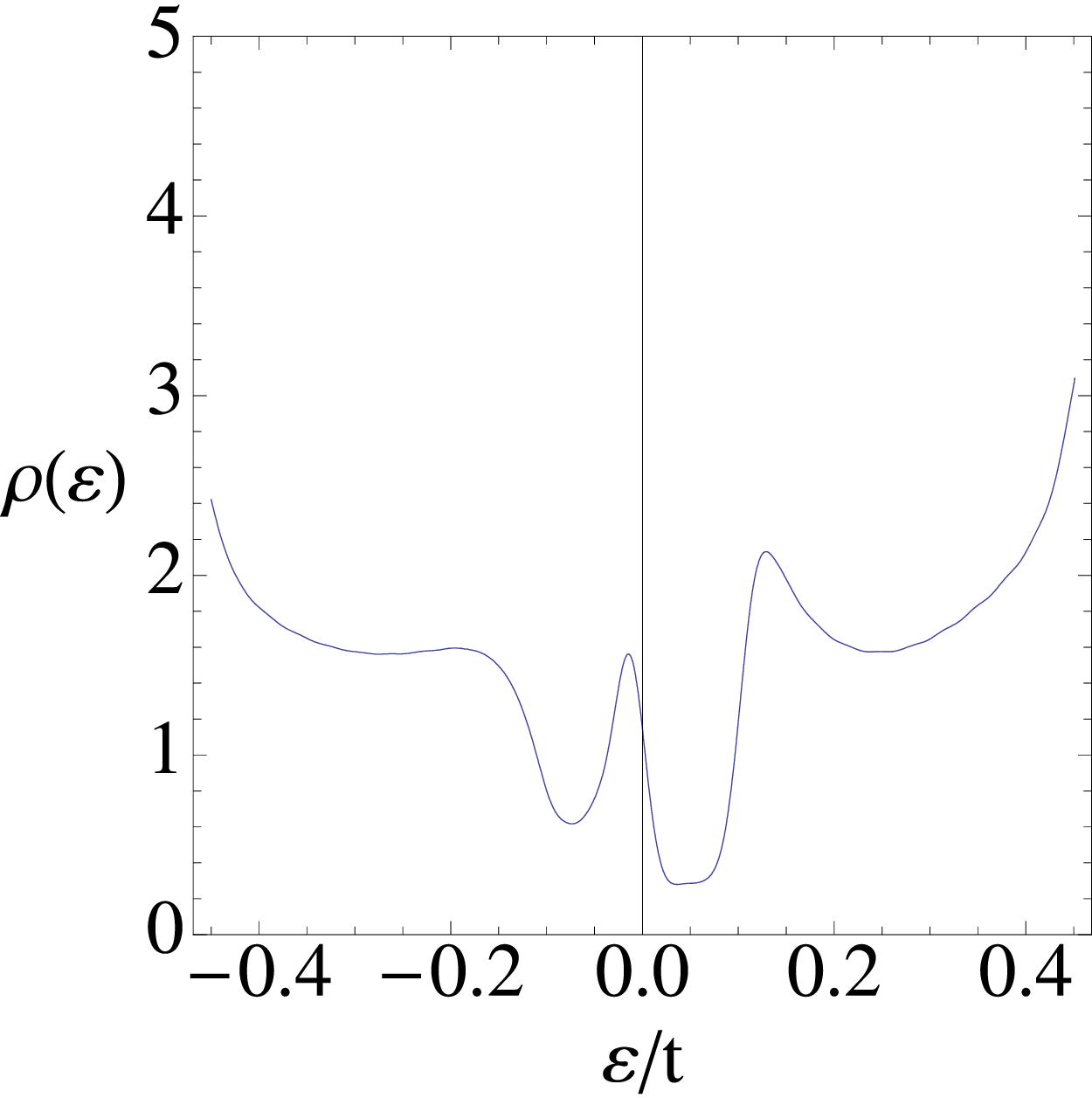}  }\\
 \hspace{0.65cm}  (a) \hspace{4.75cm} (b)   \\   \mbox{}  \\
\mbox{
\epsfxsize 4cm \epsfbox{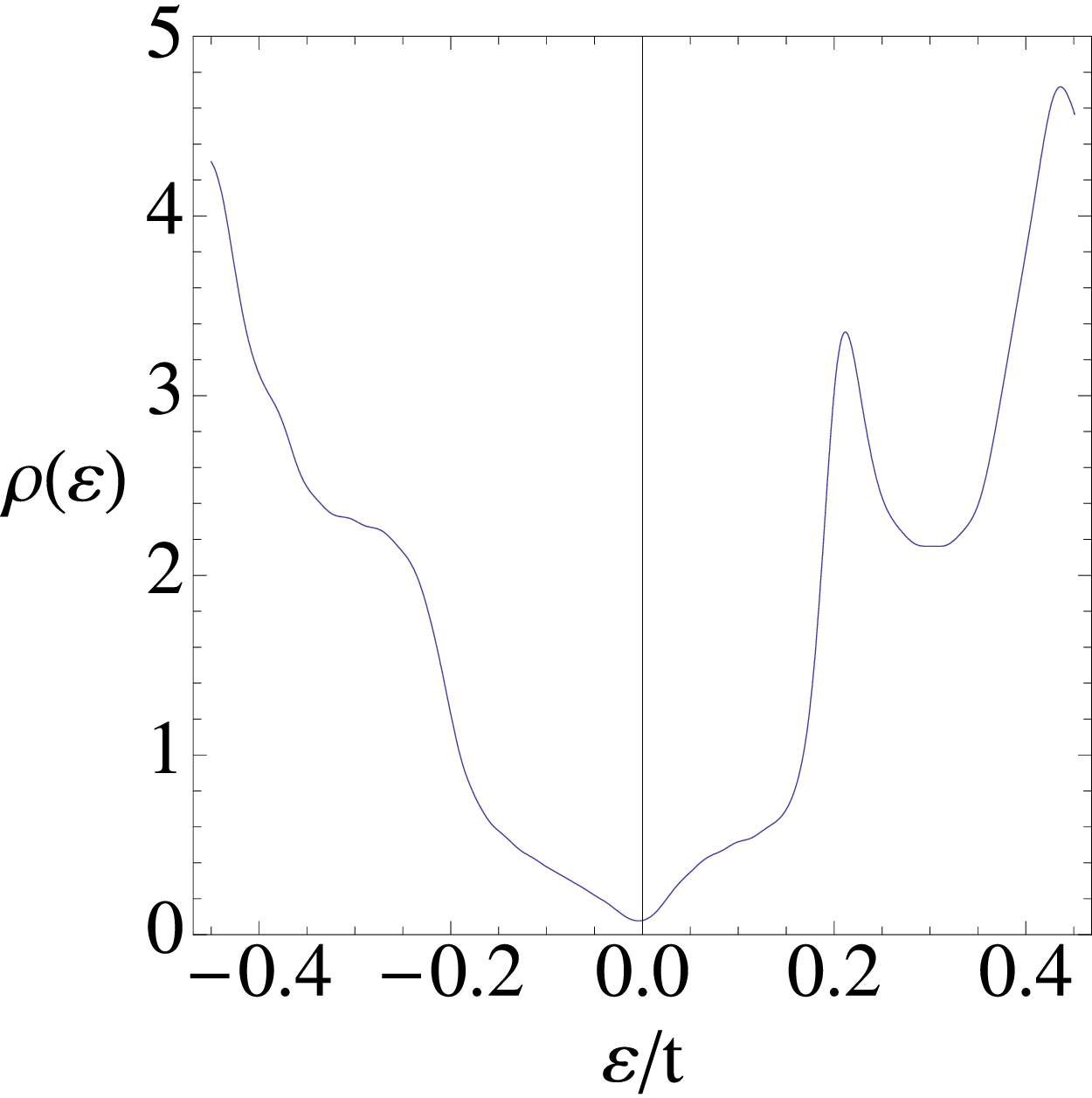}  \hspace{0.5cm}
\epsfxsize 4cm     \epsfbox{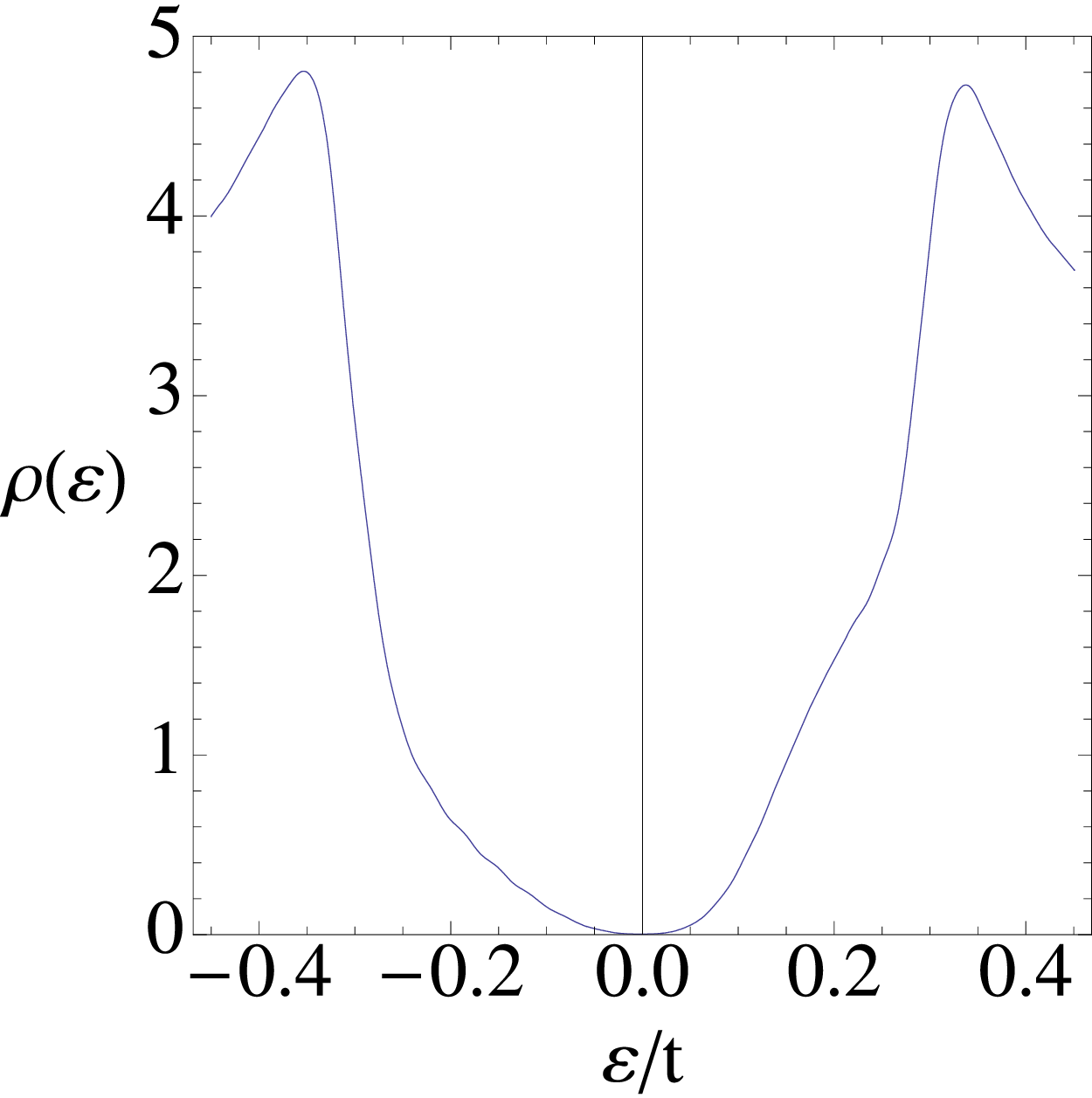} }\\
 \hspace{0.65cm}  (c) \hspace{4.75cm} (d)
\end{center}
\caption{Sequence of local densities of states for a circular ring
of atoms at the end of a $(54,0)$ nanotube close to the junction,
for the different sectors corresponding to eigenvalue $q$ under
$\pi /3$ rotation equal to 1 (a), $e^{\pm i\pi /3}$ (b),
$e^{\pm 2i\pi /3}$ (c), and $-1$ (d). Energy is measured in units of the
transfer integral $t$.}
\label{two}
\end{figure}

Our computations have covered a number of structures including armchair and
zig-zag nanotubes with different radii. After inspection of all the spectra,
it becomes clear that there are several generic features in the density of
states. We have represented in Fig. \ref{two} the behavior near the junction
between a graphene layer and a (54,0) zig-zag nanotube. We observe that,
apart from the peak close to zero energy in the sectors corresponding to
$q = e^{\pm i\pi /3}$, for $q \neq 1$ there is always a depletion in the
density of states at low energies, delimited by two abrupt upturns. It is
remarkable that the pattern in the sectors corresponding to
$q = e^{\pm 2i\pi /3}$ reproduces the same observed for $q = e^{\pm i\pi /3}$,
but with a scale that is
approximately twice larger. The density of states in the sector with
$q = -1$ displays in turn a wider depletion, with the position of the
peaks scaled by an approximate factor of 3 with respect to those in the
$q = e^{\pm i\pi /3}$ sectors.

\begin{figure}
\begin{center}
\mbox{
\epsfxsize 4cm \epsfbox{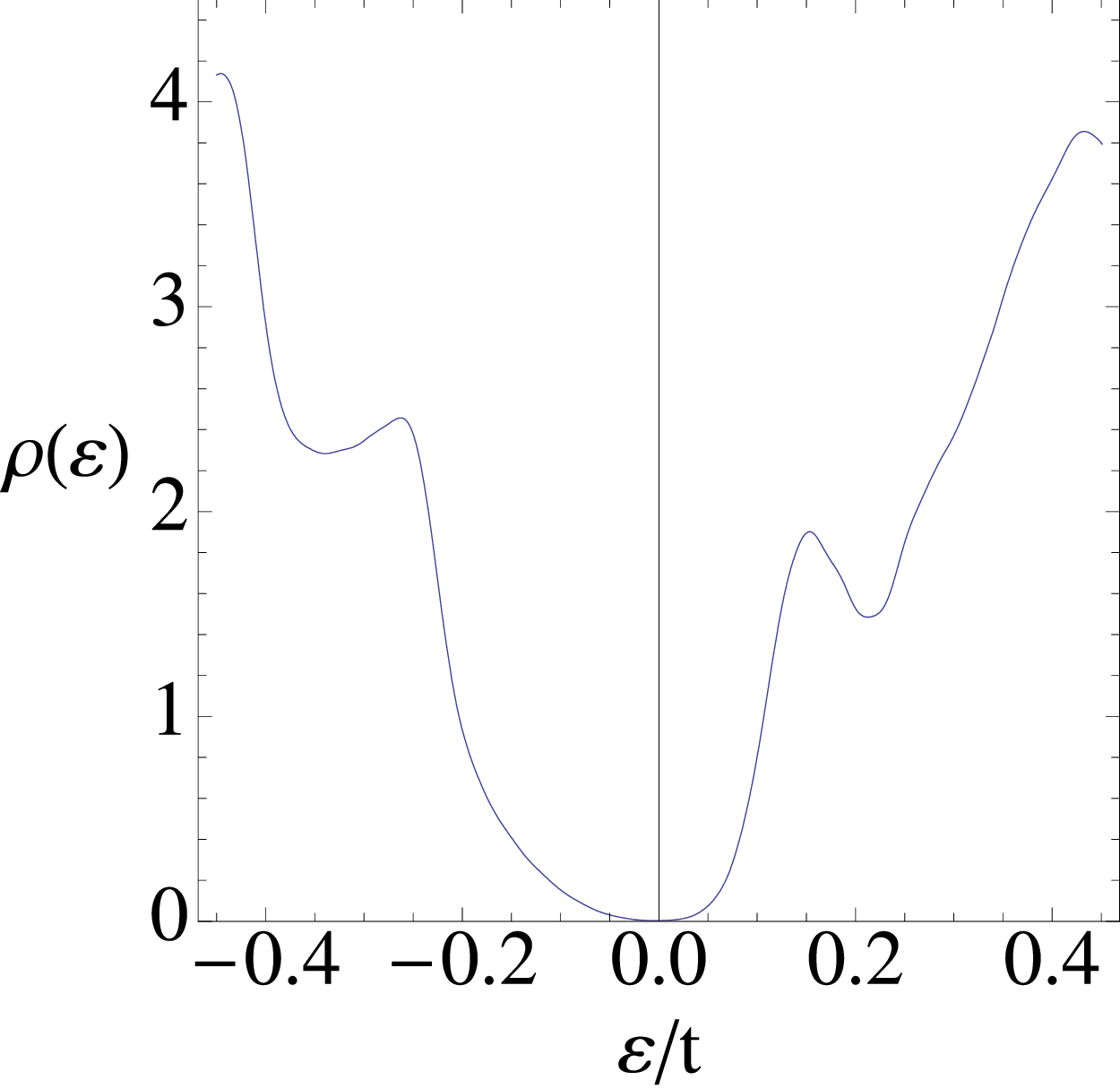}   \hspace{0.5cm}
\epsfxsize 4cm   \epsfbox{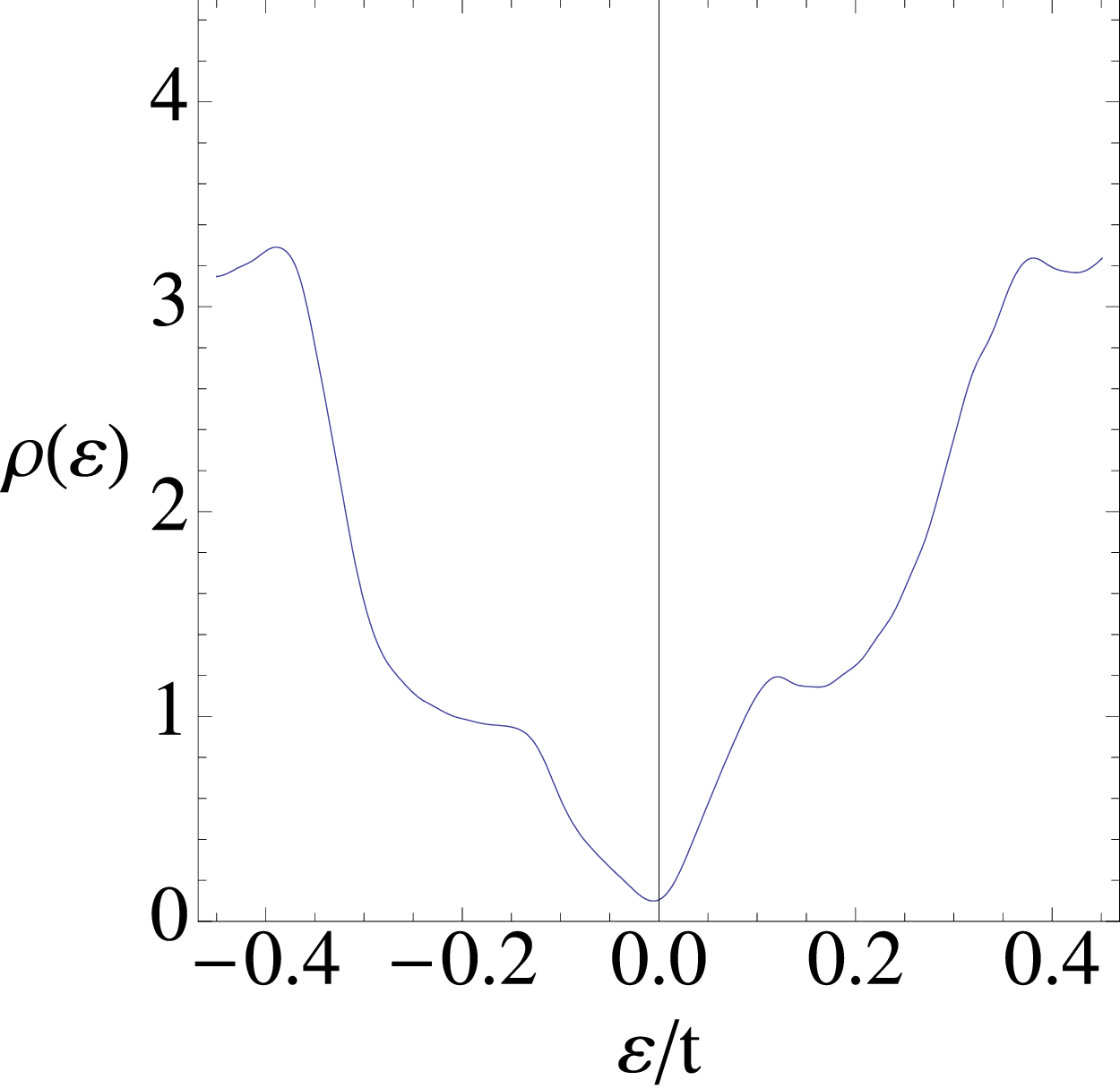}  }\\
 \hspace{0.65cm}  (a) \hspace{4.75cm} (b)   \\   \mbox{}  \\
\mbox{
\epsfxsize 4cm \epsfbox{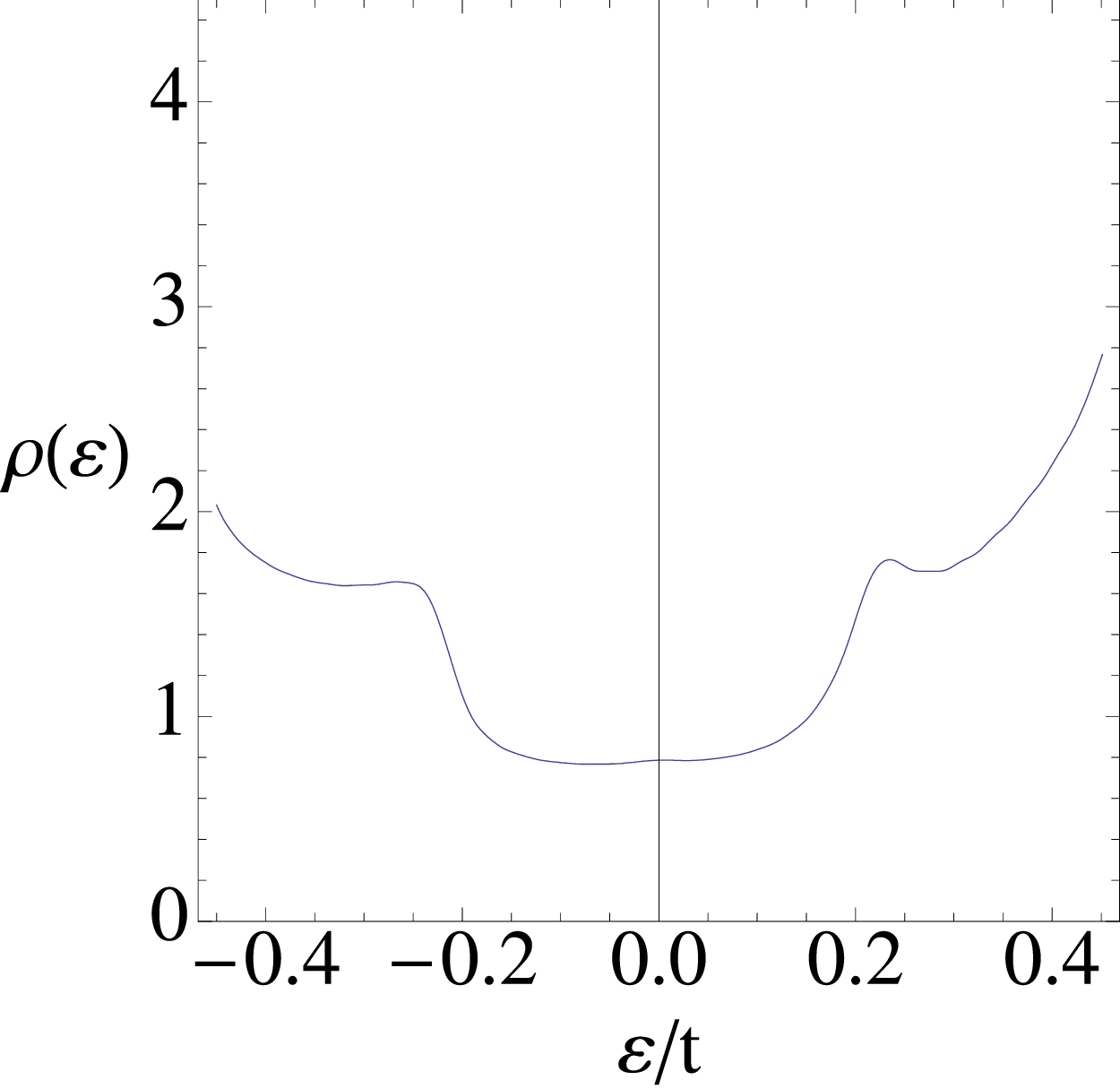}   \hspace{0.5cm}
\epsfxsize 4cm    \epsfbox{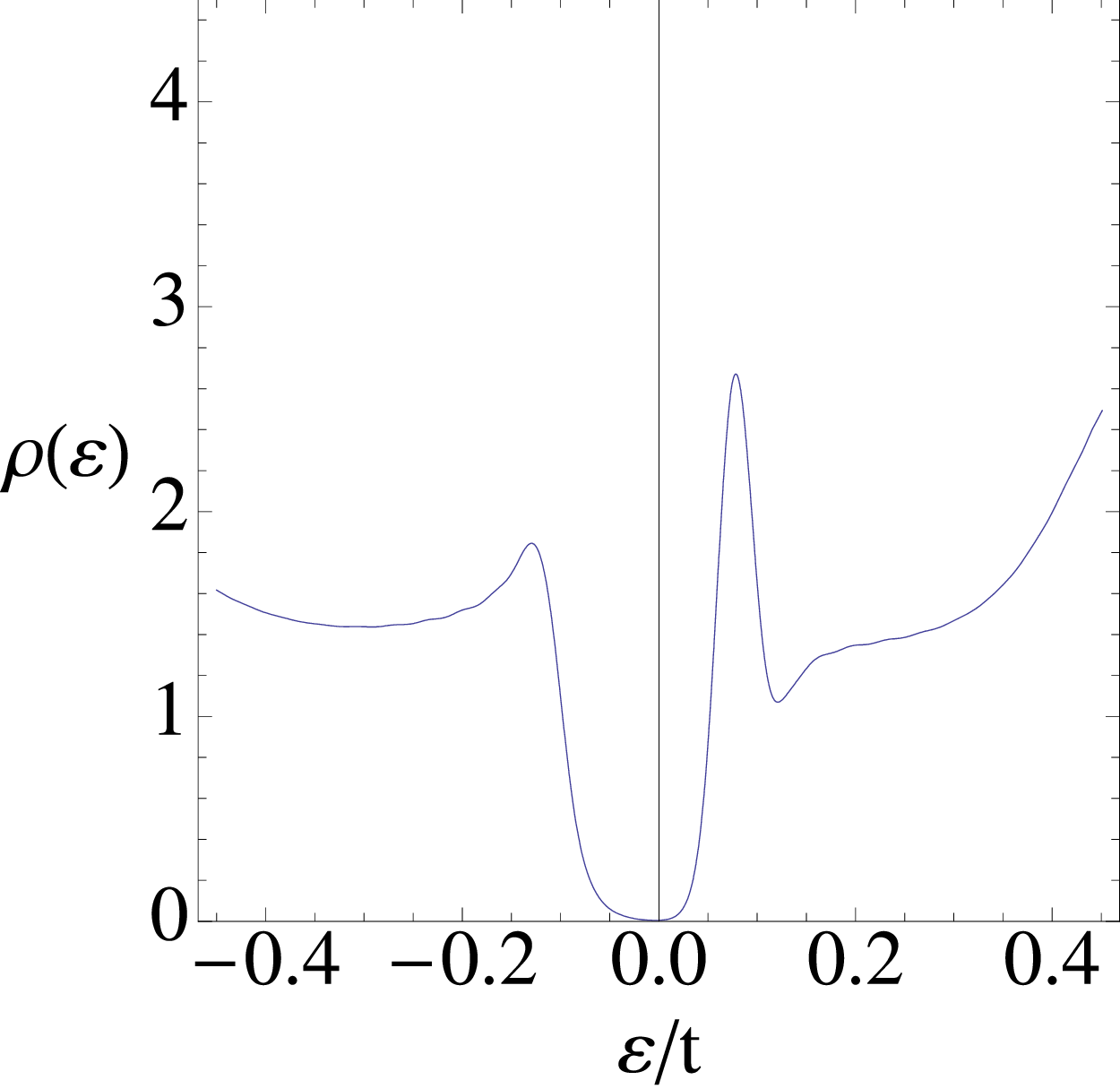} }\\
 \hspace{0.65cm}  (c) \hspace{4.75cm} (d)
\end{center}
\caption{Similar sequence as in Fig. \ref{two} for a $(48,0)$ nanotube close
to the junction, for $q = 1$ (a), $e^{\pm i\pi /3}$ (b),
$e^{\pm 2i\pi /3}$ (c), and $-1$ (d).}
\label{three}
\end{figure}

In the above behavior of the density of states, the appearance of the
peak close to zero in the sectors with $q = e^{\pm i\pi /3}$ is the only
feature not generic for all kinds of nanotubes. In the junctions made with
zig-zag nanotubes, the peak actually appears for nanotube geometries of the
type $(6n, 0)$ when $n$ is a multiple of 3. In this series of hybrid
structures, the patterns in the density of states for each value of $q$ are
quite similar, with the position of the corresponding peaks scaled in
proportion to the radius of the nanotube. On the other hand, the rest of
junctions, for which $n$ is not a multiple of 3, display a different behavior.
We have represented in Fig. \ref{three} the local density of states averaged
over a ring of atoms at the end of a $(48,0)$ nanotube close to the junction.
It can be observed the depletion of the density of states at low
energies in all but one of the $q$-sectors, and the absence of a peak at
zero energy in any of the sectors.

The local density of states of the $(48,0)$ nanotube is dominated
at the junction by contributions from states with $q = e^{\pm 2i\pi /3}$,
and this has to do with the fact that the
lowest-energy subbands of a $(6n, 0)$ zig-zag nanotube have a nonvanishing
angular momentum equal to $\pm 4n$ for the motion around the tubule. This
corresponds to a quantum number $q = e^{\pm 2i \pi /3}$ in the case of the
$(48,0)$ nanotube, while the low-energy states have $q = 1$ in the $(54,0)$
nanotube. The present picture becomes then consistent with the fact that
states in higher subbands may propagate across the junction only above
(or below) some threshold energy. This feature will be established more
precisely in the continuum approach derived below in terms of the Dirac
equation.

At this point, the hybrid structures can be classified into two different
groups, depending on whether there is a peak or not close to zero energy in
the local density of states around the nanotube-graphene junction.
The peak comes actually from the contribution of a doubly degenerated level
with states having $q = e^{\pm i\pi /3}$, and whose probability distribution
decays exponentially in the nanotube. The character of these states will be
established in the next section, after developing the continuum limit in
terms of Dirac fermion fields.

\begin{figure}
\begin{center}
\mbox{
\epsfxsize 4cm \epsfbox{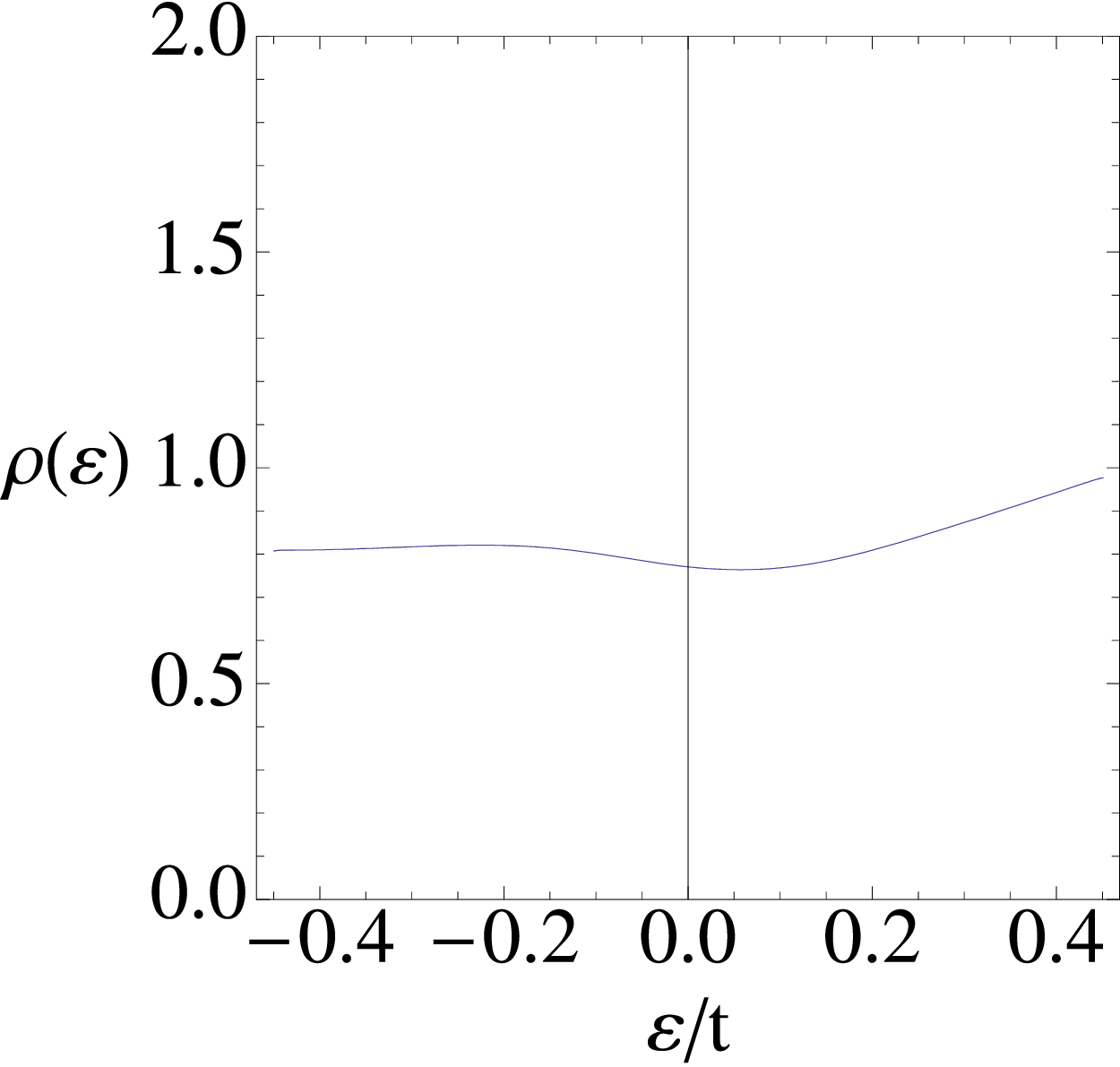}   \hspace{0.5cm}
\epsfxsize 4cm   \epsfbox{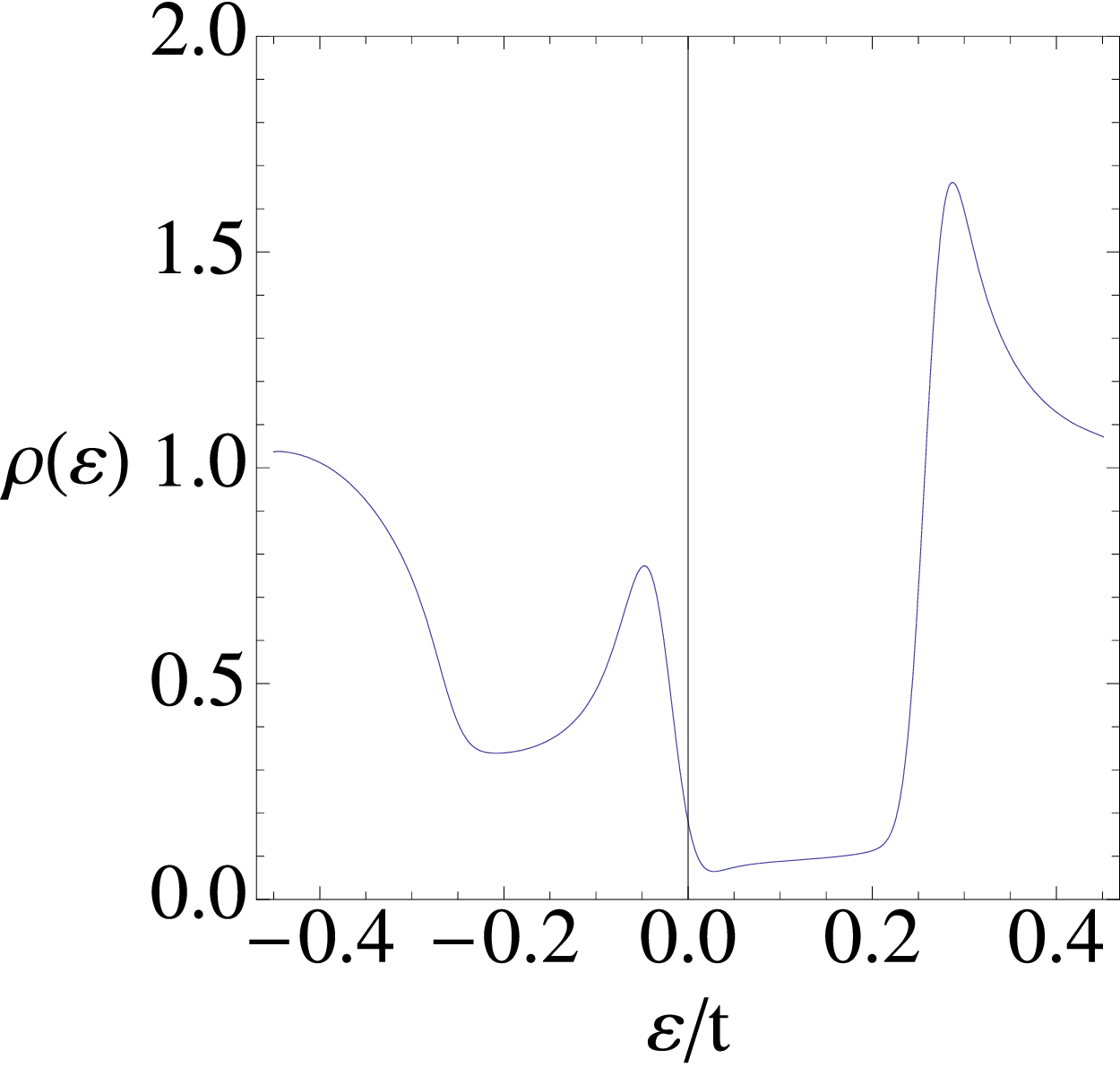}  }\\
 \hspace{0.65cm}  (a) \hspace{4.75cm} (b)   \\   \mbox{}  \\
\mbox{
\epsfxsize 4cm \epsfbox{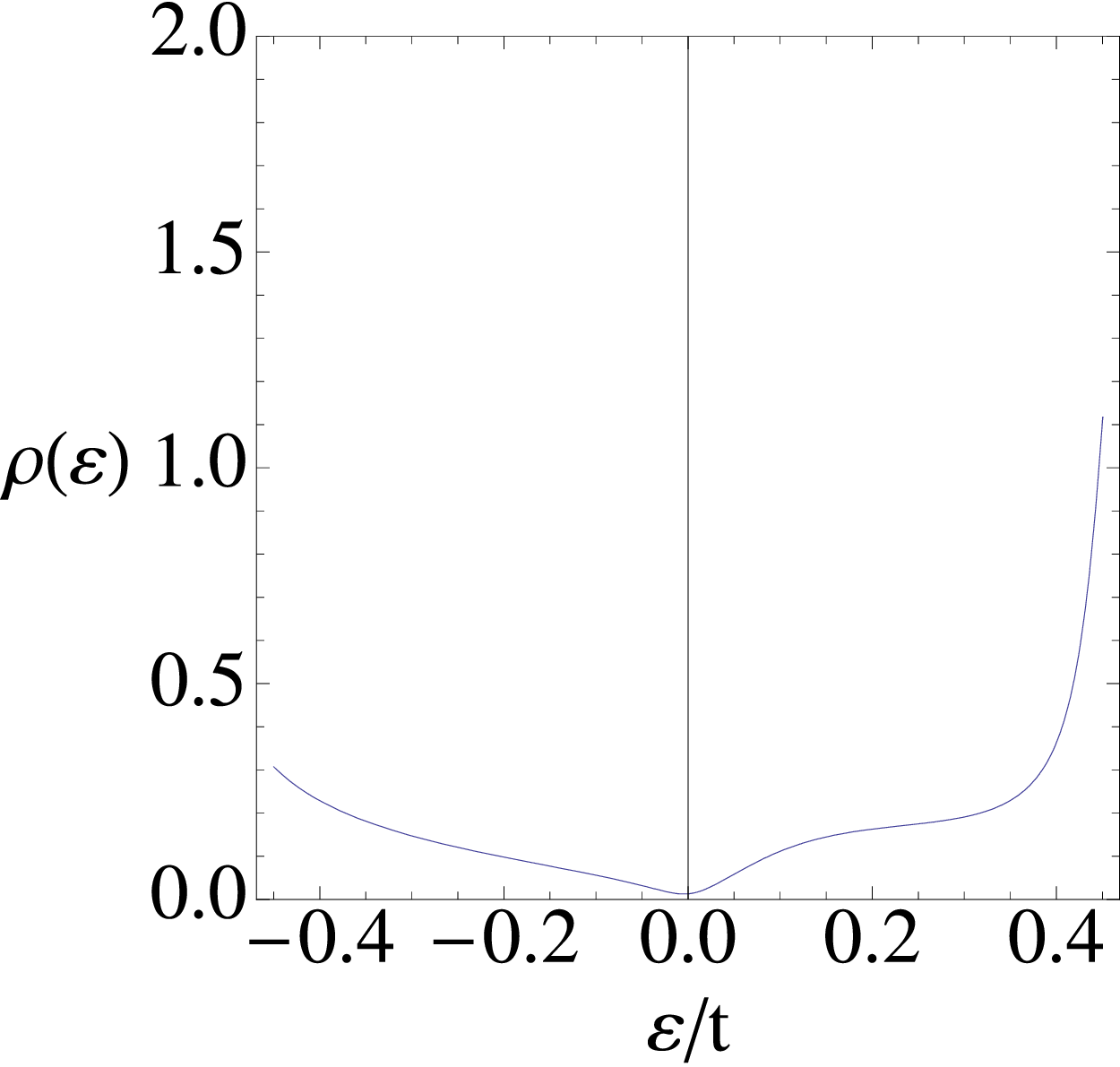}   \hspace{0.5cm}
\epsfxsize 4cm    \epsfbox{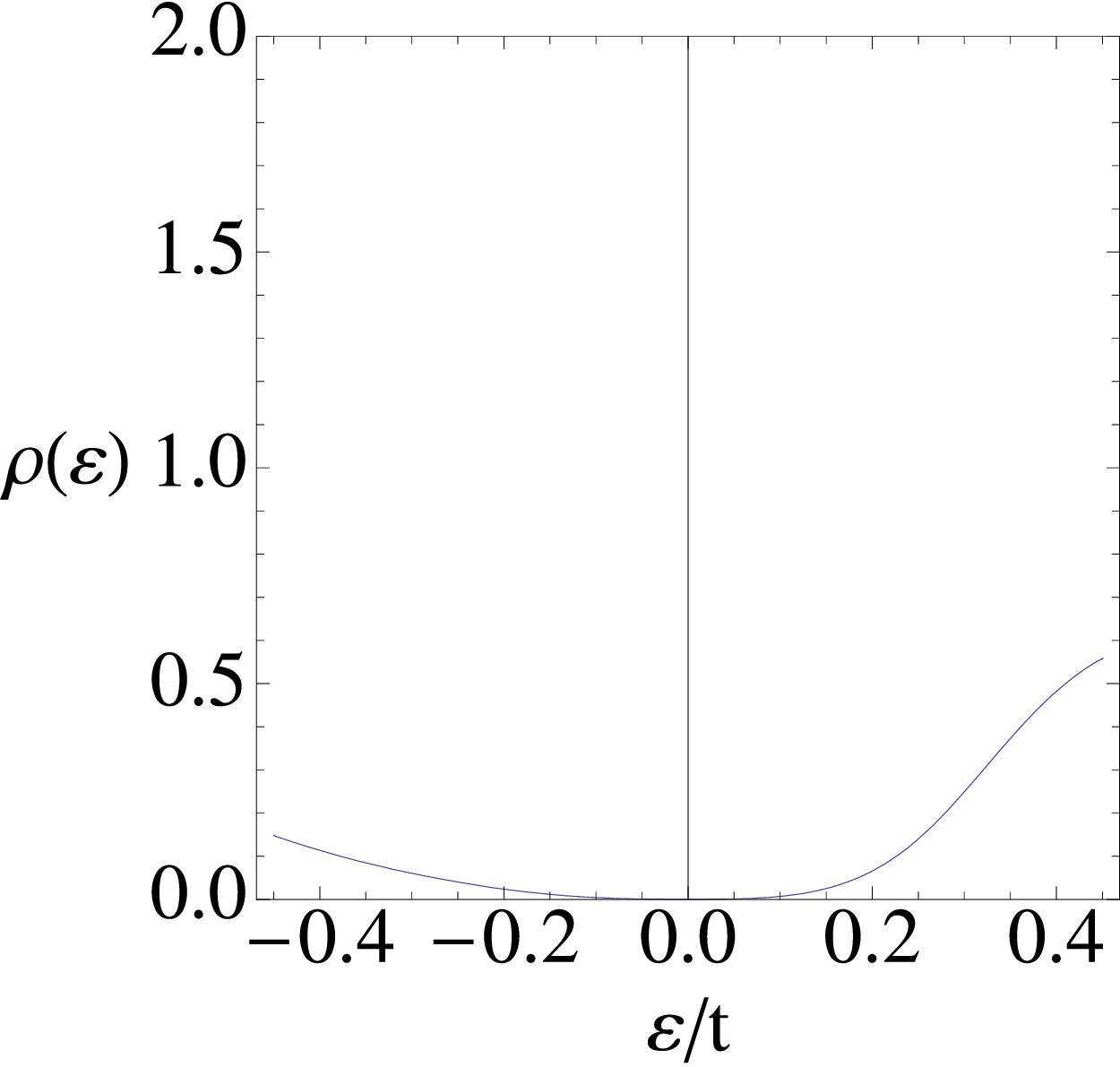} }\\
 \hspace{0.65cm}  (c) \hspace{4.75cm} (d)
\end{center}
\caption{Similar sequence as in Fig. \ref{two} for a $(12,12)$ nanotube close
to the junction, for $q = 1$ (a), $e^{\pm i\pi /3}$ (b),
$e^{\pm 2i\pi /3}$ (c), and $-1$ (d).}
\label{four}
\end{figure}

The evidence for the two different classes of hybrid structures is
reinforced by the fact that the junctions made with armchair nanotubes
behave in a quite similar way to that shown by the $(6n, 0)$ nanotubes when
$n$ is a multiple of 3. The local density of states averaged over a circular ring
of atoms around the junction between a $(12,12)$ nanotube and a graphene layer
has been represented in Fig. \ref{four}. We observe the presence of
the peak close to zero energy in the sectors with $q = e^{\pm i\pi /3}$. There
is a clear depletion in the local density of states at low energies except
in the sector with $q = 1$, which is consistent with the fact that the
lowest-energy subbands in the armchair nanotube correspond to zero
angular momentum around the tubule.

\begin{figure}
\begin{center}
\mbox{
\epsfxsize 4cm \epsfbox{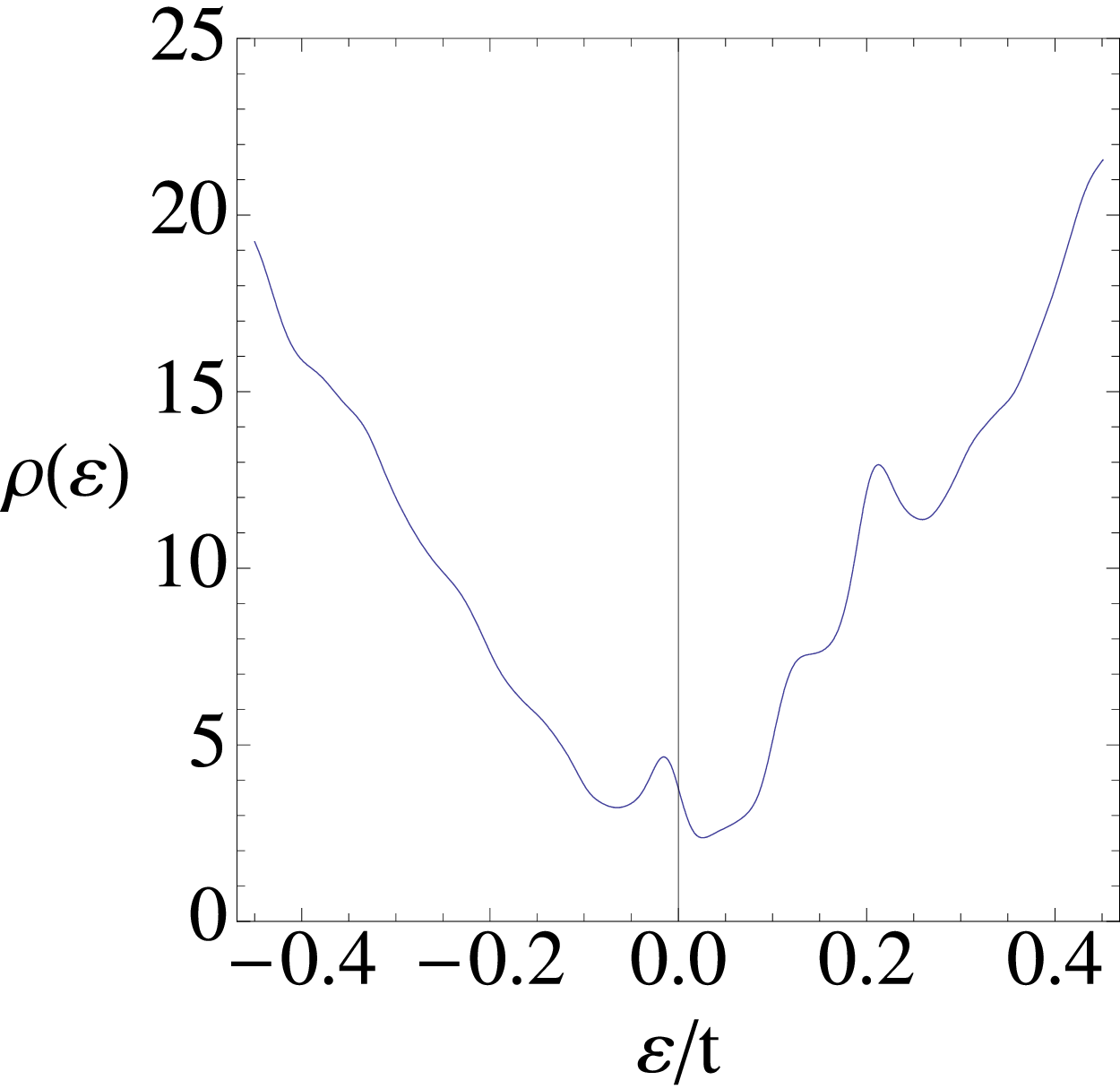}   \hspace{0.5cm}
\epsfxsize 4cm   \epsfbox{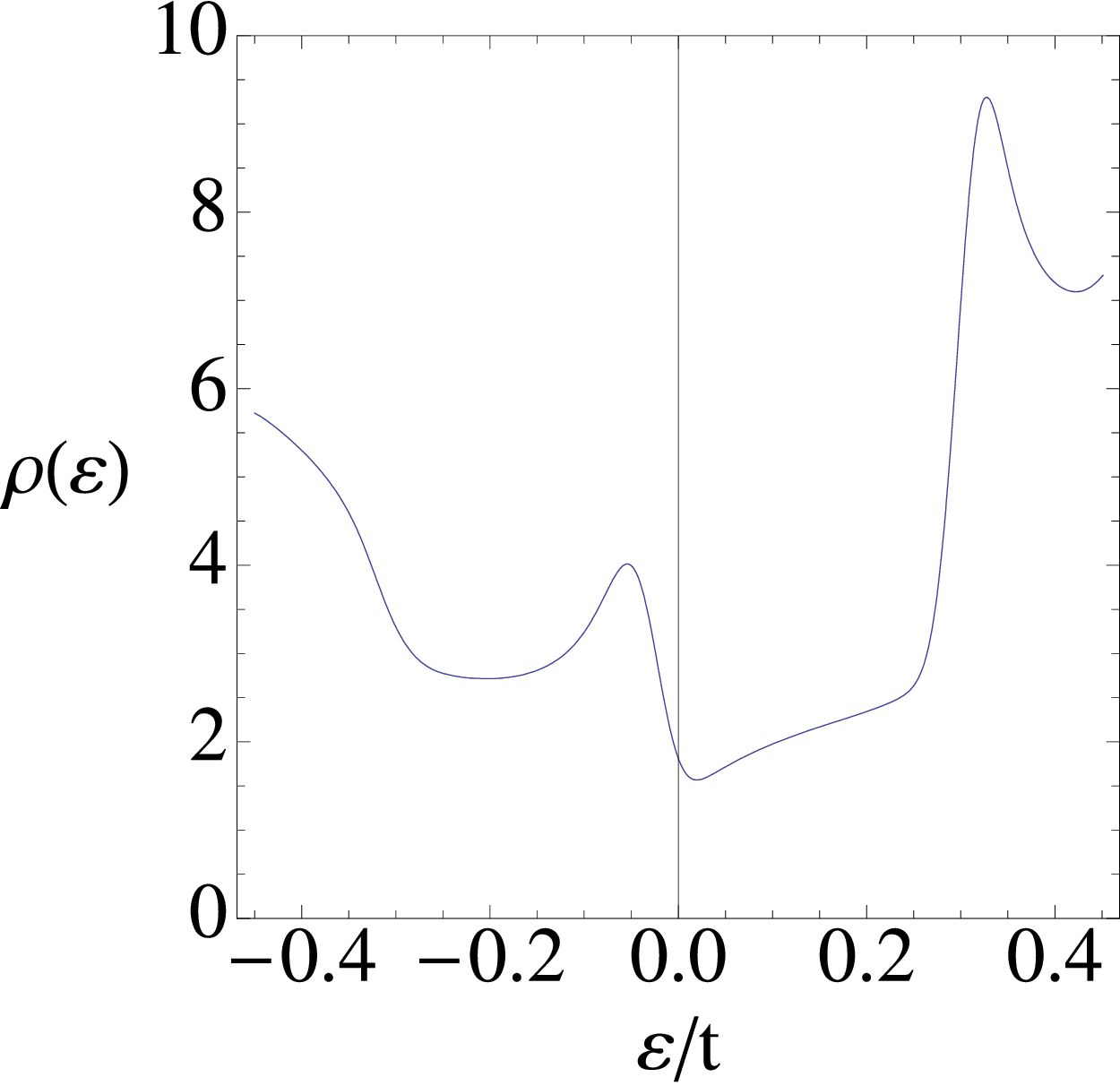}  }\\
 \hspace{0.65cm}  (a) \hspace{4.75cm} (b)   \\   \mbox{}  \\
\mbox{
\epsfxsize 4cm \epsfbox{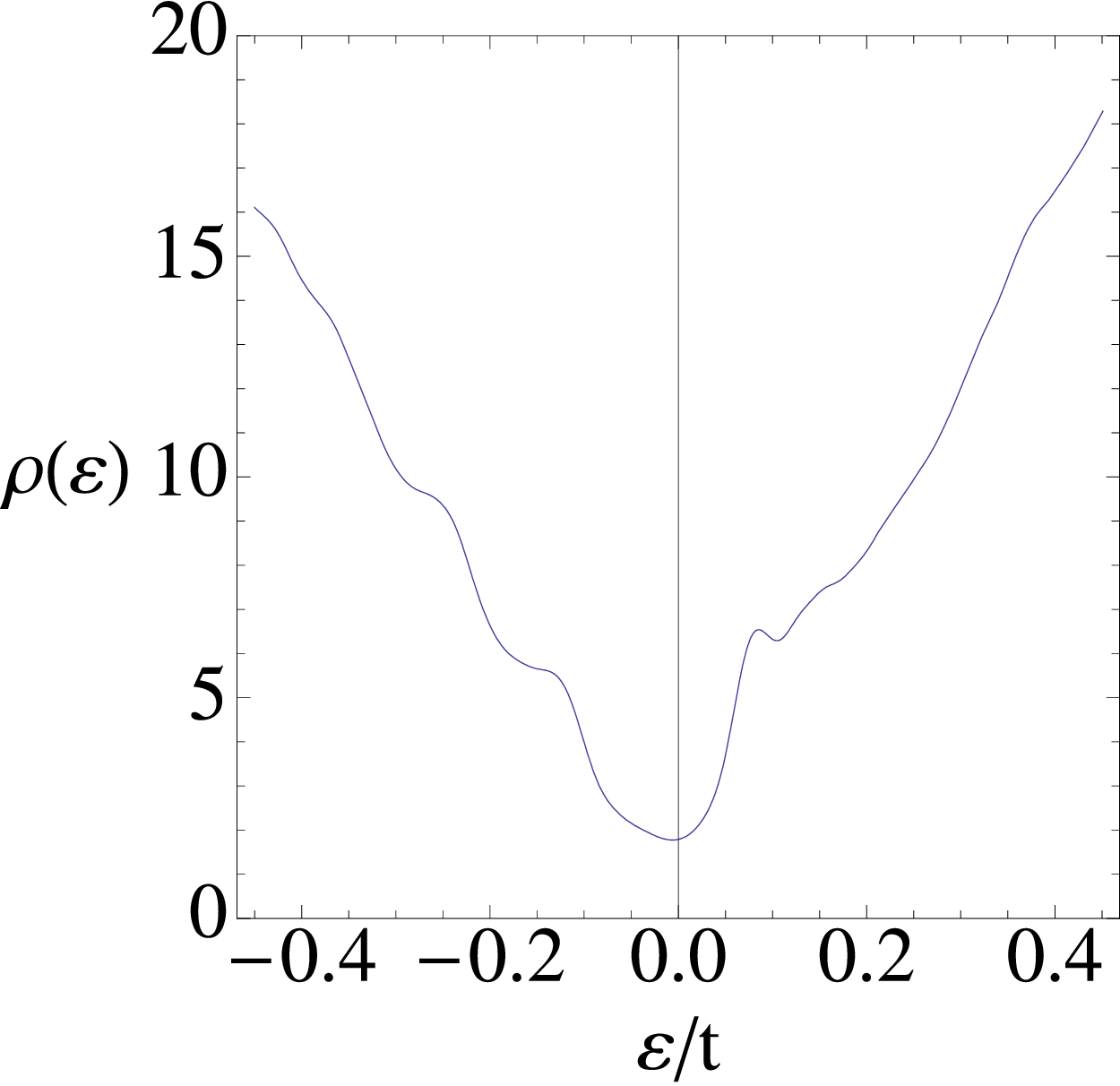}   \hspace{0.5cm}
\epsfxsize 4cm    \epsfbox{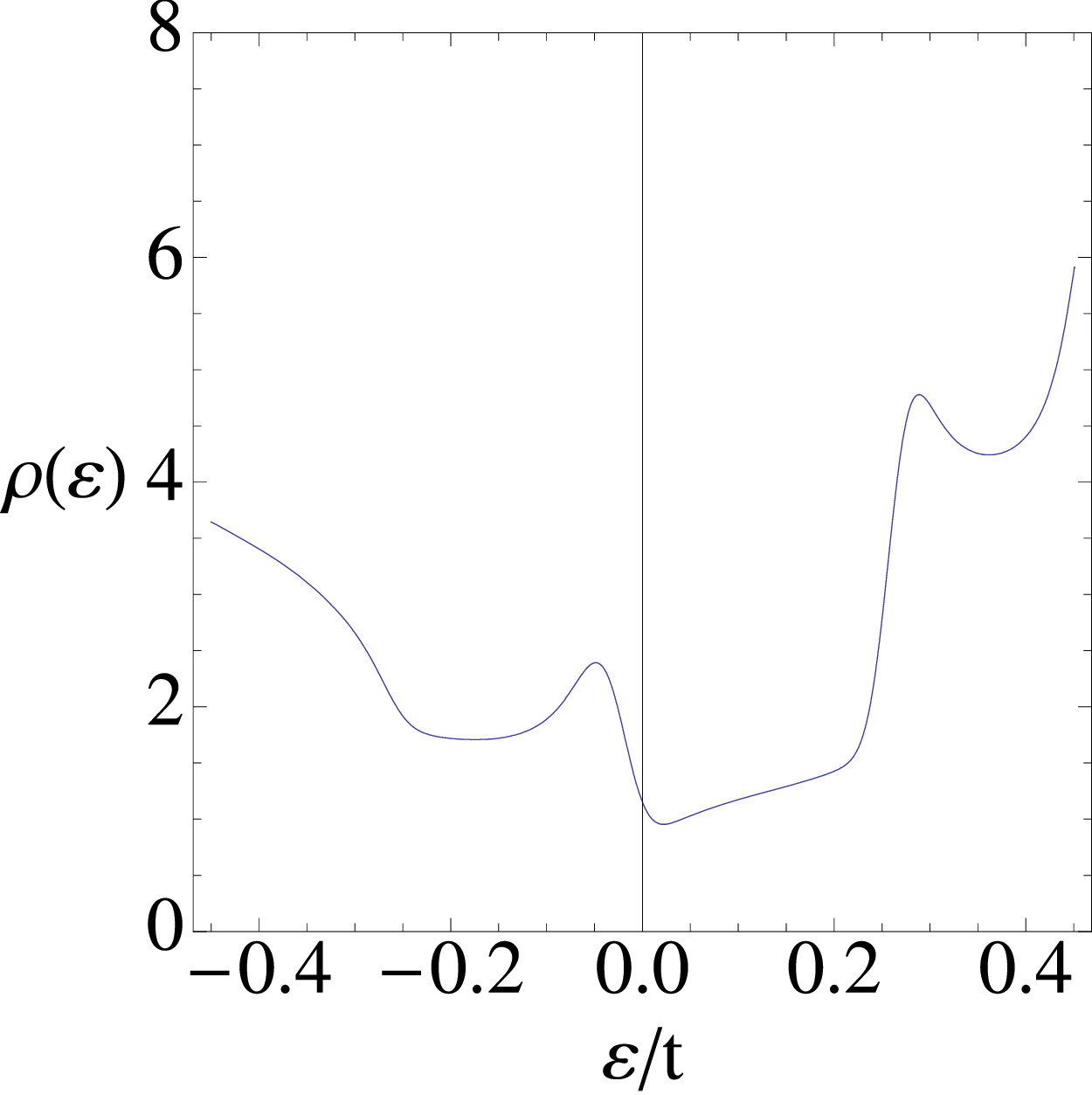} }\\
 \hspace{0.65cm}  (c) \hspace{4.75cm} (d)
\end{center}
\caption{Sequence of local densities of states for a ring
of atoms at the end of the nanotube close to the junction,
for the different geometries
$(54,0)$ (a), $(18,0)$ (b), $(48,0)$ (c), and $(12,12)$ (d).
Energy is measured in units of the transfer integral $t$.}
\label{five}
\end{figure}

The existence of the two different classes of nanotube-graphene
junctions is illustrated in Fig. \ref{five}, which shows the results for
the local density of states around the junction (after summing over the
sectors with different values of $q$) for the different types of nanotube
considered above. It is remarkable the similarity between the density of
states for the $(18,0)$ and $(12,12)$ nanotube geometries, which have a
very close value of the radius. This suggests that there must be a universal
way of understanding the low-energy electronic properties of the two
different classes of junctions, independent of the details of the lattice
building the junction within each class.

\section{Continuum approach}

\subsection{Heptagonal rings, effective flux, and matching at the junction}

We observe that the density of states summed over all values of $q$ tends 
to have an approximate linear behavior away from the very low energy regime. 
This motivates the analysis of the junction in terms of the Dirac equation 
in the hybrid geometry, as the Dirac fermions provide an appropriate 
description of the electronic properties in the 1D carbon nanotube as well 
as in the 2D graphene layer.

We will then assume that the radius $R_0 $ of the nanotube is much larger
than the graphene lattice constant, in order to obtain the continuum limit 
of the tight-binding model. Within this approximation, the analysis of
the electronic structure is reduced to the study of the Dirac equation in
a space with an abrupt change from a planar to a cylindrical
structure. The transition from a geometry to the other takes place
due to the presence of the six heptagonal rings at the junction.
These defects are the source of negative curvature, playing a role
opposite to that of the pentagonal rings in a fullerene cage\cite{ggv}.

The heptagons, as well as the pentagons, also induce frustration in the
honeycomb lattice, and lead to the exchange of the two Dirac valleys of the
planar geometry\cite{ggv2}. This latter effect has to be accounted for by means
of an effective  non-abelian gauge field operating in the space of the two
independent Dirac points of graphene. It can be shown that the effective flux
associated to an individual heptagonal ring is equal to $\pi /2 $ \cite{ggv2}
(in units such that $\hbar = 1$). The flux provided by the six heptagons at
the junction can reach therefore a maximum of 3/2 times the flux quantum. In
general, however, the count of the total flux may not follow an additive rule,
so that it can be lower than the maximum value, depending on the relative
position of the heptagonal rings\cite{lc}. We will see that this is actually
the origin of the two different classes of junctions.

Note that the angular momentum around the axis of the nanotube is conserved 
and quantized in integer units, in a continuum description of the geometry 
analyzed here. On the other hand, the angular momentum in the plane is quantized 
and shifted by $1/2$ plus the number of flux quanta of the effective gauge field 
induced by the heptagonal rings. The existence of topological defects which induce 
an effective flux at the junction allows us to match wavefunctions with different 
angular momenta at either side of the junction, provided that the effective flux
corresponds to a half-integer number of quanta.

\subsection{Continuum wavefunctions}

By looking at the effect of pairs of heptagonal rings, the non-abelian
gauge field operating in the space of the two Dirac points becomes anyhow
proportional to a sigma matrix $\tau_3 $. We may consider then that the effect
of the six heptagonal rings at the junction is described by an effective
abelian field, standing for either of the eigenvalues of $\tau_3 $ \cite{ggv2}.
We end up therefore
with two different Dirac equations, with effective magnetic fluxes of opposite
sign. We will denote the Dirac spinors satisfying the two Dirac equations as
$\Psi^{+}$ and $\Psi^{-}$, respectively. Given that the region of the space
away from the junction has no curvature, we can use radial coordinates
$r, \theta $ to write the set of two Dirac equations in the plane for
$r \geq R_0 $:
\begin{eqnarray}
i v_F \left( \partial_r + \frac{i\partial_{\theta } }{r} \pm \frac{g}{r}
         + \frac{1}{2r}  \right) \Psi_A^{\pm } (r,\theta )
   & = &  \varepsilon  \Psi_B^{\pm } (r,\theta )
\label{dp1}                                                            \\
i v_F \left( \partial_r - \frac{i\partial_{\theta } }{r} \mp \frac{g}{r}
         + \frac{1}{2r}  \right) \Psi_B^{\pm } (r,\theta )
   & = &  \varepsilon  \Psi_A^{\pm } (r,\theta )
\label{dp2}
\end{eqnarray}
$v_F$ is the Fermi velocity and the components $\Psi_A$ and $\Psi_B$ denote 
the respective amplitudes of the electron in the two sublattices of the 
graphene lattice. The parameter $g$ corresponds to the quanta of
effective flux felt by the electrons when making a complete tour around the
junction. 

In the nanotube side, we use cylindrical coordinates $z, \theta $, with
$z \leq 0$. The Dirac equation for the nanotube is
\begin{eqnarray}
 i v_F \left( \partial_z +  \frac{i\partial_{\theta } }{R_0}  \right)
  \Psi_A^{\pm } (z,\theta )  & = &  \varepsilon  \Psi_B^{\pm } (z,\theta )
\label{dn1}                                                               \\
 i v_F \left( \partial_z -  \frac{i\partial_{\theta } }{R_0}  \right)
  \Psi_B^{\pm } (z,\theta )  & = &  \varepsilon  \Psi_A^{\pm } (z,\theta )
\label{dn2}
\end{eqnarray}

We note that (\ref{dp1})-(\ref{dp2}) as well as (\ref{dn1})-(\ref{dn2})
are expressions of the Dirac equation in flat space. This is
consistent with the fact that, in the continuum limit, the curvature is
localized at the circle connecting graphene and the nanotube. As an
alternative to the coordinate $z$, we could make for instance the change of
variables $r = R_0 \exp (z/R_0 )$, allowing us to map the nanotube into
the region of the plane with $r < R_0$. Using a common radial coordinate
$r > 0$ to describe both graphene and the nanotube, the metric of
the space turns out to be multiplied by the conformal factor
$\Omega (r) = \theta (r - R_0 ) + (R_0 / r)^2 \theta (R_0 - r ) $. The first
derivative of the metric becomes then discontinuous and the curvature scalar,
computed in terms of second derivatives of the metric, is
$R = - 2r \delta (r - R_0)$. The integral of this expression corresponds
actually to the total curvature provided by the heptagonal rings. This makes
clear that the effects of the curvature are implicit in the operation of
matching the solutions of (\ref{dp1})-(\ref{dp2}) and (\ref{dn1})-(\ref{dn2})
at the circle $r = R_0$.

The solutions of the Eqs. (\ref{dp1})-(\ref{dp2}) are of the form
\begin{equation}
\left(
\begin{array}{c}
\Psi_A^{\pm } (r,\theta )  \\
\Psi_B^{\pm } (r,\theta )
\end{array}
\right)    \equiv   c_1
\left(
\begin{array}{c}
J_{n \mp g - \frac{1}{2}} (kr)  \\
-i \: \text{sgn} (\varepsilon ) \: J_{n \mp g + \frac{1}{2}} (kr)
\end{array}
\right)
e^{i n \theta }   +   c_2
\left(
\begin{array}{c}
Y_{n \mp g - \frac{1}{2}} (kr)  \\
-i \: \text{sgn} (\varepsilon ) \: Y_{n \mp g + \frac{1}{2}} (kr)
\end{array}
\right) e^{i n \theta } \label{e0}
\end{equation}
where $c_1$ and $c_2$ are constants, and $J_n (x)$ and $Y_n (x)$ are Bessel
functions. The energy is $\varepsilon = \pm v_F k$.

On the other hand, the resolution of Eqs. (\ref{dn1})-(\ref{dn2}) shows that
there are propagating and evanescent waves in the nanotube, which can be
written as:
\begin{equation}
\left(
\begin{array}{c}
\Psi_A^{\pm } (z,\theta )  \\
\Psi_B^{\pm } (z,\theta )
\end{array}
\right)    \equiv
\left\{
\begin{array}{ll}
c'_1
\left(
\begin{array}{c}
1   \\
- \text{sgn} (\varepsilon ) \: e^{i \phi (k)}
\end{array}
\right)
e^{ik z}  e^{in \theta }  +  c'_2
\left(
\begin{array}{c}
1   \\
 \text{sgn} (\varepsilon ) \: e^{-i \phi (k)}
\end{array}
\right)
e^{-ik z}  e^{in \theta }   &  |\varepsilon | > \frac{v_F |n|}{R_0}    \\
c
\left(
\begin{array}{c}
1   \\
- \text{sgn} (\varepsilon ) \: e^{i \phi (- i \kappa)}
\end{array}
\right)
e^{\kappa z}  e^{in \theta }   &  |\varepsilon | \leq \frac{v_F |n|}{R_0}
\end{array}
\right.
\label{pe}
\end{equation}
where $c'_1, c'_2 $ and $c$ are constants, the energy $\varepsilon $ is given
by:
\begin{equation}
\varepsilon =
\left\{
\begin{array}{ll}
\pm v_F \sqrt{k^2 + \frac{n^2}{R_0^2}}  &
                            |\varepsilon | > \frac{v_F |n|}{R_0}    \\
\pm v_F \sqrt{- \kappa^2 + \frac{n^2}{R_0^2}}  &
                            |\varepsilon | \leq \frac{v_F |n|}{R_0}
\end{array}
\right.
\label{e}
\end{equation}
and the phase factor in Eq. (\ref{pe}) is:
\begin{equation}
e^{i \phi (k)} = \frac{k + \frac{in}{R_0}}{\sqrt{k^2 + \frac{n^2}{R_0^2}}}
\end{equation}

We note that the evanescent states with longitudinal decay $e^{\kappa z}$
arise for nonvanishing angular momentum $n$. Then there is an energy threshold
$v_F |n|/R_0 $ for the appearance of propagating states in the nanotube. This
is perfectly consistent with the behavior of the local density of states
obtained for the different values of $q$ in the tight-binding approach. The
depletion found in different $q$-sectors for the local density of states at
the end of the nanotube (close to the junction) corresponds actually to the
range of evanescent states given by (\ref{e}). As already mentioned, the
position of the peaks delimiting the depletion in the tight-binding approach
scales in proportion to the value of the angular momentum, which corresponds 
in the lattice to the different values of $q$. Moreover, we have also seen
that such a position is inversely proportional to the nanotube radius $R_0 $,
with values in the plots that can be approximately matched with the estimate
$v_F n/R_0 $ (after using the expression of the Fermi velocity $v_F =
3 t a /2$, in terms of the transfer integral $t$ and the C-C distance $a$).
We find therefore that the generic features found for the local density of
states in the tight-binding approach are well captured by the continuum limit
based on the Dirac equation.

\subsection{Zero-energy states}

The presence of a peak in the local density of states close to zero energy
(in the sectors $q = e^{\pm i\pi /3 }$) is the only feature not generic for all
types of nanotubes, and that can be also explained within our continuum
approach. The rotation caused by each heptagonal ring in the space of the two
Dirac points corresponds to an effective magnetic flux of $\pi /2 $ \cite{ggv2},
but the way this flux is combined in the case of pairs of heptagons depends on
their relative position. This has been studied in the case of pentagon
pairs in Ref. \onlinecite{lc}, arriving at a conclusion that can be readily
generalized to the case of heptagonal defects. The result is that,
when the distance between the heptagons is given by a vector $(N,M)$ (using
the same notation to classify the nanotubes) such that $N-M$ is not a
multiple of 3, the effective flux of a pair of heptagons does not add to
$\pi $, but to the lower amount $\pi /3$. The number $N-M$ for the distance
between heptagons is a multiple of 3 only in the case of junctions
with armchair nanotubes, or with $(6n,0)$ nanotubes when $n$ is a multiple of
3. In these instances, the total flux felt around the junction is equal to the
sum of the fluxes provided by the individual heptagons, giving a value of 
$g = 3/2$. In the rest of the cases, the total flux corresponds instead to 
$g = 1/2$.

The number $g$ of flux quanta has a direct correspondence with the number
of zero modes of the Dirac equation. Their existence rests on the possibility
of having localized states at the junction, with suitable decay in both
the graphene part and the nanotube side. If we take for instance the maximum
effective flux and $g = 3/2$, we have an equation for a zero-energy eigenstate
in the region $r > R_0$
\begin{equation}
\left( \partial_r + \frac{i\partial_{\theta }}{r} + \frac{2}{r} \right)
 \Psi_A (r,\theta ) = 0
\end{equation}
For a wavefunction with angular momentum $n$, we obtain the behavior
\begin{equation}
\Psi_A (r,\theta ) \sim r^{n-2} e^{in \theta }
\label{zm}
\end{equation}
This gives rise to modes decaying from the junction for values
$n \leq 1 $  . The wavefunction has to be matched at $r = R_0$
with the appropriate dependence in the nanotube, that is
\begin{equation}
\Psi_A (z,\theta ) \sim  e^{\frac{n}{R_0} z}  e^{in \theta }
\end{equation}
Recalling that $z \leq 0$, we see that only the value $n = 1$ provides a
localized state at the junction. On the graphene side, the state is not
strictly normalizable, in a similar way to other half-bound states induced by
defects\cite{pgl,dhm}. Anyhow, such a localized state has a reflection in the
peak observed close to zero energy in the $q = e^{\pm i\pi /3}$ sectors of the
tight-binding density of states. By inverting the direction of the flux and
taking $g = -3/2$, it can be seen that the solutions have then a nonvanishing
component $\Psi_B (r,\theta )$ similar to (\ref{zm}), but with angular
momentum $-n$ instead of $n$. Another localized state is found therefore with
opposite chirality and $n = -1$.

In the case of the junctions with $(6n,0)$ nanotubes such that $n$ is not a
multiple of 3, the flux corresponding to $g = 1/2$ is not enough to localize
states at the junction. It can be seen that there are no zero-energy solutions
of the Dirac equation decaying simultaneously in the graphene plane and in the
nanotube. This explains why in this type of junctions there is no low-energy
peak within the depleted region of the local density of states. We complete
in this way the correspondence between the tight-binding approach and the
continuum limit based on the Dirac equation, accounting  for the main
electronic features and unveiling also the origin of the different low-energy
behavior in the two classes of junctions.

\subsection{Transmission into the nanotube}

Eqs. (\ref{e0}) and (\ref{pe}) allow us to analyze the scattering of
a wave in the plane off the nanotube. The coefficients $c_1 , c_2 ,
c'_1$ and $c'_2$ define the transmission and reflection by the
nanotube. For a wave coming from the plane, the coefficient $c'_2$
is zero, and $c'_1$ is proportional to the transmission coefficient.
Using the theory of scattering of two dimensional Dirac electrons by
an impurity\cite{OGM06,hg,KN07,N07,G08}, the transmission
coefficient is given by:
\begin{equation}
T_n = \sqrt{\frac{\pi k R_0}{2}} \frac{i J_{n+1} ( k R_0 ) Y_n ( k R_0 ) -
i J_n ( k R_0 ) Y_{n+1} ( k R_0 )}{i Y_n ( k R_0 ) - Y_{n+1} ( k R_0 ) e^{i
\phi (k)}}
\end{equation}
At high energies, $\epsilon \gg v_F / R_0$ or, alternatively, $k R_0
\rightarrow \infty$, we can use the asymptotic expansion:
\begin{eqnarray}
\lim_{k R_0 \rightarrow \infty} J_n ( k R_0 ) &\approx
&\sqrt{\frac{2}{\pi k R_0}} \cos \left( k R_0 - \frac{n \pi}{2} -
\frac{\pi}{4} \right)
\nonumber \\
\lim_{k R_0 \rightarrow \infty} Y_n ( k R_0 ) &\approx
&\sqrt{\frac{2}{\pi k R_0}} \sin \left( k R_0 - \frac{n \pi}{2} -
\frac{\pi}{4} \right)
\end{eqnarray}
and $\lim_{k R_0 \rightarrow \infty} e^{i \phi (k)} = 1$. From these
expansions, we obtain:
\begin{equation}
\lim_{k R_0 \rightarrow \infty} T_n \approx \frac{i}{i \sin \left( k R_0
- \frac{n \pi}{2} - \frac{\pi}{4} \right) + \cos \left( k R_0 -
\frac{n \pi}{2} - \frac{\pi}{4} \right) e^{i \phi (k)}} \rightarrow
e^{i \left( k R_0 - \frac{n \pi}{2} - \frac{\pi}{4} \right)}
\end{equation}
so that $\lim_{k R_0 \rightarrow \infty} | T_n |^2 = 1$. This estimate
is valid for angular momenta $n$ such that $n \ll k R_0$. We can also
obtain the reflection coefficient in this limit, $\bar{R}_n$, which
is also independent of $n$ for $n \ll k R_0$. The angular dependence
of the scattering cross section $\sigma ( \theta )$ is:
\begin{equation} \sigma
( \theta ) \propto \frac{1}{k} \left[ \frac{\sin ( k R_0 \theta
)}{\sin ( \theta / 2 )} \right]^2
\end{equation}
and the total cross section, $\sigma = \int \sigma ( \theta ) d
\theta$, is proportional to $R_0$. The total flux of particles
propagating inside the nanotube, normalized to the total incoming
flux, is also proportional to $R_0$.

In the low energy limit, $k R_0 \ll 1$, we obtain:
\begin{equation}
\lim_{k R_0 \rightarrow 0} | T_n |^2 \approx \left\{ \begin{array}{lr}
0 &k R_0 \le n-1 \\ \frac{\pi}{n !^2} \left( \frac{k R_0}{2}
\right)^{2n+1} &k R_0 > n-1 \end{array} \right.
\end{equation}
In this limit, most of the electrons reaching the junction are
scattered back into the plane.

\subsection{Local density of states}

We can make use of the continuum equations to analyze systems of
very large sizes.  We calculate the electronic Green's functions
numerically. The Dirac equation in the plane, in radial coordinates,
can be discretized\cite{WSG07}. Each radial equation can be
approximated by a nearest neighbor tight binding model with two
inequivalent hoppings:
\begin{eqnarray}
t \left( 1 + \frac{2 n + 1}{4 i} \right) a_i + t \left ( 1 - \frac{2
n + 1}{4 i} \right) a_{i+1} &=  &\epsilon b_i \nonumber \\
t \left( 1 + \frac{2 n + 1}{4 i} \right) b_{i-1} + t \left( 1 -
\frac{2 n + 1}{4 i} \right) b_i &= & \epsilon a_i \label{discrete}
\end{eqnarray}
where $a_i$ and $b_i$ give the values of $\Psi_A^n ( r )$ and
$\Psi_B^n ( r )$ at position $r = i \times a$, $a$ being a length
scale which defines the discretization. Eqs. (\ref{discrete}) also
include an energy scale, $t$, which plays the role of an upper
cutoff. The Fermi velocity is equal to $t \times a$. The Dirac equation
is obtained for $\epsilon \ll t$. We can write Eqs. (\ref{discrete})
in a more compact form using a single index, $c_i$, such that $c_{2 j
- 1} = a_j$ and $c_{2j} = b_j$, so that:
\begin{equation}
t_i c_{i-1} + t_{i+1} c_{i+1} = \epsilon c_i
\end{equation}
where, using as new length scale $a/2$, we have:
\begin{equation}
t_i = t \left[ 1 - (-1)^i \frac{2n+1}{4 ( i + N_0 )} \right]
\end{equation}
where we start at position $N_0$. The diagonal Green's function at
site $i$ can be written as:
\begin{equation}
{\cal G}_{ii} ( \epsilon ) = \frac{1}{\epsilon - t_i T_+^i - t_{i+1}
T_-^{i+1}} \label{green}
\end{equation}
and:
\begin{eqnarray}
T_+^{i+1} &= &\frac{t_{i+1}}{\epsilon - t_i T_+^i} \nonumber \\
T_-^i &= &\frac{t_i}{\epsilon - t_{i+1} T_-^{i+1}}
\end{eqnarray}
with boundary conditions at $i=0$ and $i=N$:
\begin{eqnarray}
T_+^0 &= &\frac{1}{\epsilon - \Sigma_0 ( \epsilon )} \nonumber \\
T_-^N &= &\frac{1}{\epsilon - \Sigma_N ( \epsilon )}
\end{eqnarray}
and:
\begin{eqnarray}
\Sigma_0 ( \epsilon ) &= &\frac{t^2 \left( 1 + \frac{n-1}{2 N_0}
\right)^2}{\epsilon} + \frac{\tilde{\epsilon}}{2} -
\frac{\sqrt{\tilde{\epsilon}^2 - 4 \tilde{t}^2}}{2} \nonumber \\
\Sigma_N ( \epsilon ) &= &i t \label{bc}
\end{eqnarray}
and:
\begin{eqnarray}
\tilde{\epsilon} &= &\epsilon - \frac{t^2}{\epsilon} \left( 1 +
\frac{(n-1)^2}{4 N_0^2} \right) \nonumber \\
\tilde{t} &= &\frac{t^2}{\epsilon} \left( 1 - \frac{(n-1)^2}{4
N_0^2} \right)
\end{eqnarray}
where the radius of the nanotube is $R_0 = N_0 \times a / 2$. The
boundary conditions in Eq. (\ref{bc}) describe a semi-infinite
nanotube attached at position $N_0$, and also approximate the
boundary of a plane at position $N$.
\begin{figure}
\begin{center}
\includegraphics*[width=8cm]{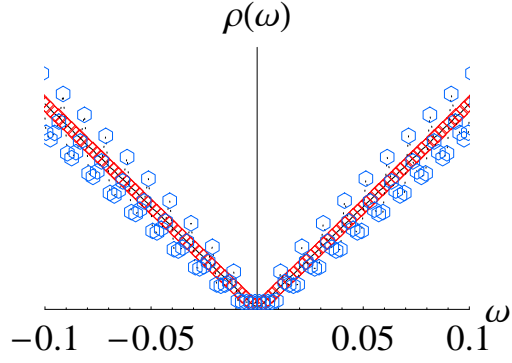}
\end{center}
\caption{Density of states at the bulk of the nanotube (blue,
hexagons), and at the graphene plane (red, diamonds), obtained using
the numerical method discussed in the text.} \label{dos_0}
\end{figure}
The Green's function deep inside the nanotube can be calculated
analytically:
\begin{equation}
{\cal G}_{ii}^{nt} ( \epsilon ) = \sum_{n = - \infty}^{n= \infty}
\frac{1}{\sqrt{\tilde{\epsilon}^2 - 4 \tilde{t}^2}} \theta \left( |
\epsilon | - \frac{v_F n}{R_0} \right)
\end{equation}
The density of states in the plane has been calculated numerically,
with $N = 1600$ and summing angular momenta from $n=-100$ to
$n=100$. The calculation is equivalent to analyzing a cluster with
$( 2 \times 100 + 1 ) \times 1600 = 321600$ sites. Results for the
Green's function in the nanotube, and at the position $i=100$, are
shown in Fig. \ref{dos_0}. The energy scale is set by $t=1$.
\begin{figure}
\begin{center}
\includegraphics*[width=8cm]{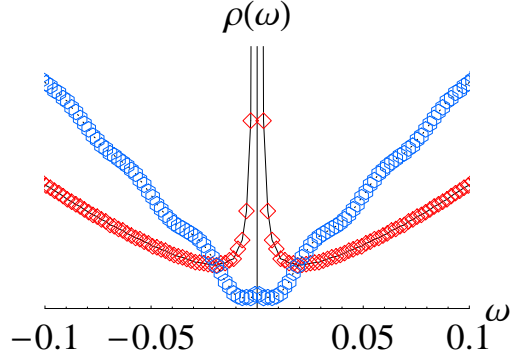}
\end{center}
\caption{Density of states at a graphene plane attached to a
nanotube. The radius of the nanotube is $N_0 = 100$. The density of
states is shown at position $n=10$ from the junction (red,
diamonds), and $n=100$ (blue, hexagons).} \label{dos}
\end{figure}

The Green's function for the system built up by the nanotube and the
graphene sheet is shown in Fig. \ref{dos}. The parameters are the
same as the ones used for the calculation shown in
Fig. \ref{dos_0}, and the radius of the nanotube is $N_0=100$.

The density of states at distances from the nanotube $n \gtrsim N_0$
are similar to those in the unperturbed sheet. Near the juncture
with the nanotube, there is a depletion of states at low energies,
compensated by the existence of a localized state at $\epsilon = 0$.

\subsection{Gauge fields due to elastic strains}

We have not analyzed so far the effect on the electronic structure
of strains which may be induced near the junction. These strains
deform the bonds, and induce an additional, intravalley gauge field
acting on the electrons\cite{Metal06,M07,NGPNG08,GKV08}. The large
in-plane stiffness of graphene implies that the bonds will tend to
their equilibrium lengths throughout the system.

The bending at the junction will be localized within a length scale
$l \sim \sqrt{\kappa / \bar{\lambda} }$, where $\kappa \sim 1$ eV is
the bending rigidity of graphene, and $\bar{\lambda} \sim 10$ eV
\AA$^{-2}$ is an average of the Lam\'e coefficients of graphene.
This length is comparable to the lattice spacing.

The mismatch between the diameter of the nanotube and the lattice
constant of the graphene layer induces additional strains, with a
long range decay into the bulk of the graphene plane and the
nanotube, which can be calculated using the continuum theory of
elasticity\cite{LL59}. We expect, however, the interatomic distance
in graphene to be very close to the distance between carbon atoms
along the radial direction of the nanotube, so that the strains
induced by this effect will be small. The strains will decay as
$r^{-1}$ or $z^{-1}$ as function of the distance to the junction.

Using dimensional analysis, the strain near the junction is of order
$\Delta R_0 / R_0$, where $\Delta R_0$ is the change in the equilibrium
radius of the nanotube induced by the plane or, alternatively, of
order $\Delta a / a$, where $a$ is the interatomic distance. We
expect the value of $\Delta a / a$ to be similar on the plane side
of the junction. The associated gauge field is $A \sim \beta \Delta
a / a^2$, where $\beta \sim \partial \log ( t ) /
\partial \log (a) \sim 2-3$ gives the change of the tight-binding
hopping $t$ with $a$.  Thus, we expect that the elastic strains will
induce changes on the electronic structure on energy scales of order
$v_F A$ near the junction.

\section{Low-energy bands in arrays of nanotube-graphene junctions}

Our computational framework allows us also to address the electronic properties
of arrays of nanotube-graphene junctions. We consider the case in which
the unit cell of the array has a hexagonal shape in the base, of the type shown 
in Fig. \ref{one}. The periodic arrangement of junctions is formed then by 
translating the unit cell by two independent vectors of the triangular array, 
in such a way that the 2D base is fully covered with the hexagonal patches. The 
Brillouin zone of the superlattice is a hexagon, and the main electronic 
properties are encoded in the form of the bands from the center to the $M$ and
$K$ points at the boundary of the zone. As long as the states in momentum
space have well-defined transformation properties under translations by the 
lattice vectors of the triangular array, we can obtain the bands of the array
of junctions by solving a tight-binding model in the unit cell, with 
appropriate momentum-dependent boundary conditions between opposite sides of 
its hexagonal base.

\begin{figure}
\begin{center}
\mbox{
\epsfxsize 4.75cm \epsfbox{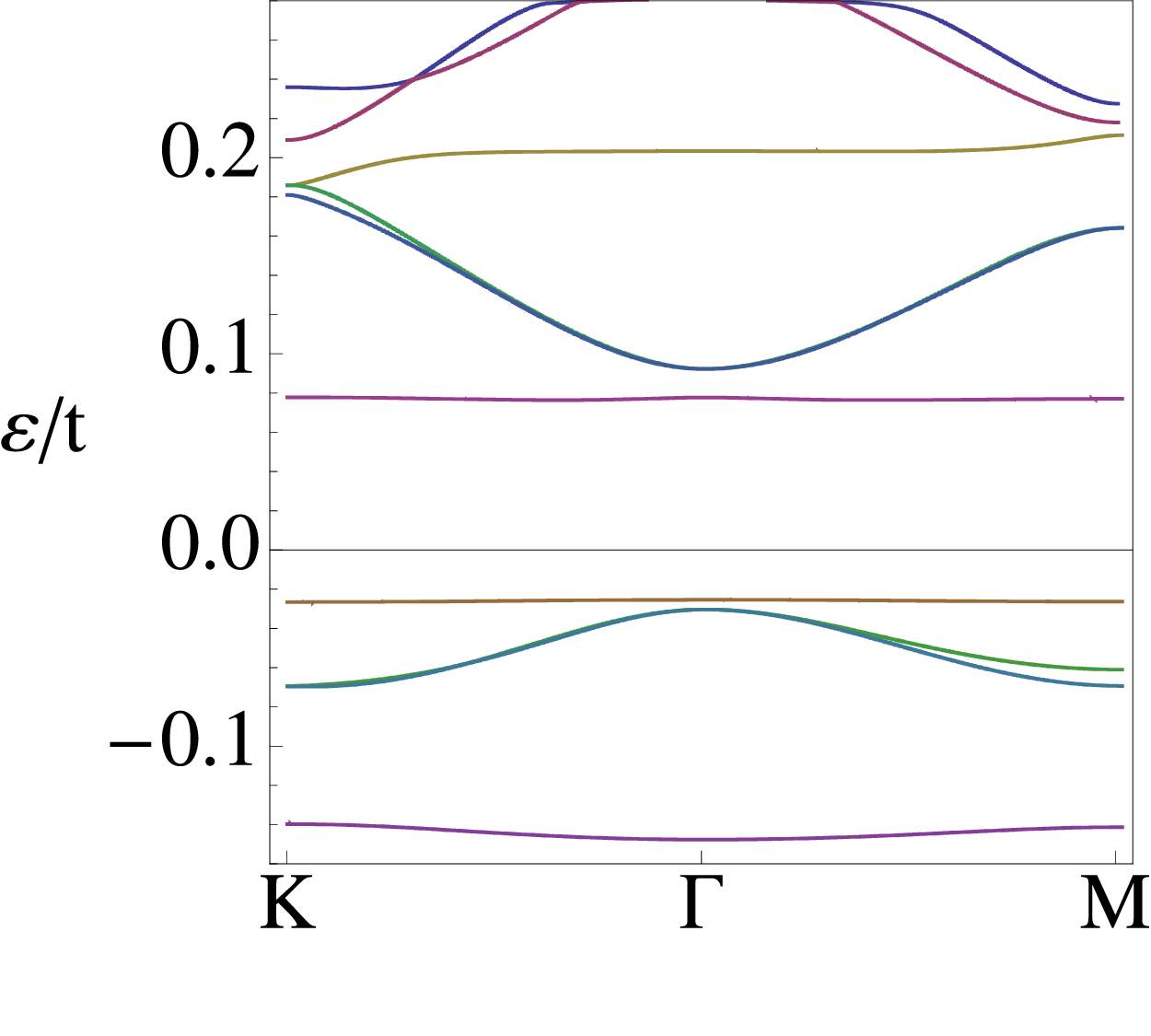}   \hspace{0.5cm}
\epsfxsize 5cm   \epsfbox{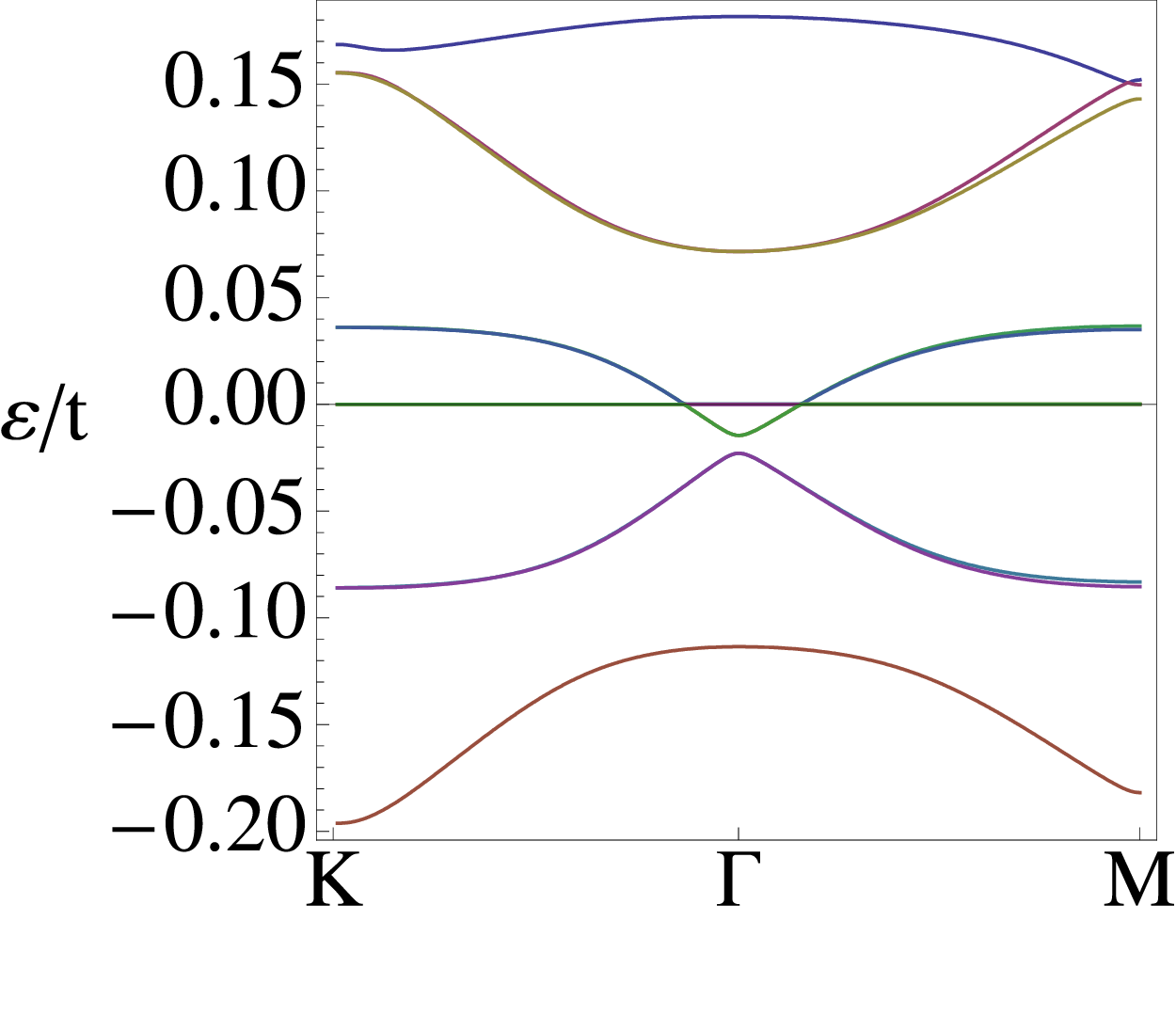}  }\\
 \hspace{0.65cm}  (a) \hspace{4.75cm} (b)   \\   \mbox{}  \\
\mbox{
\epsfxsize 4.75cm \epsfbox{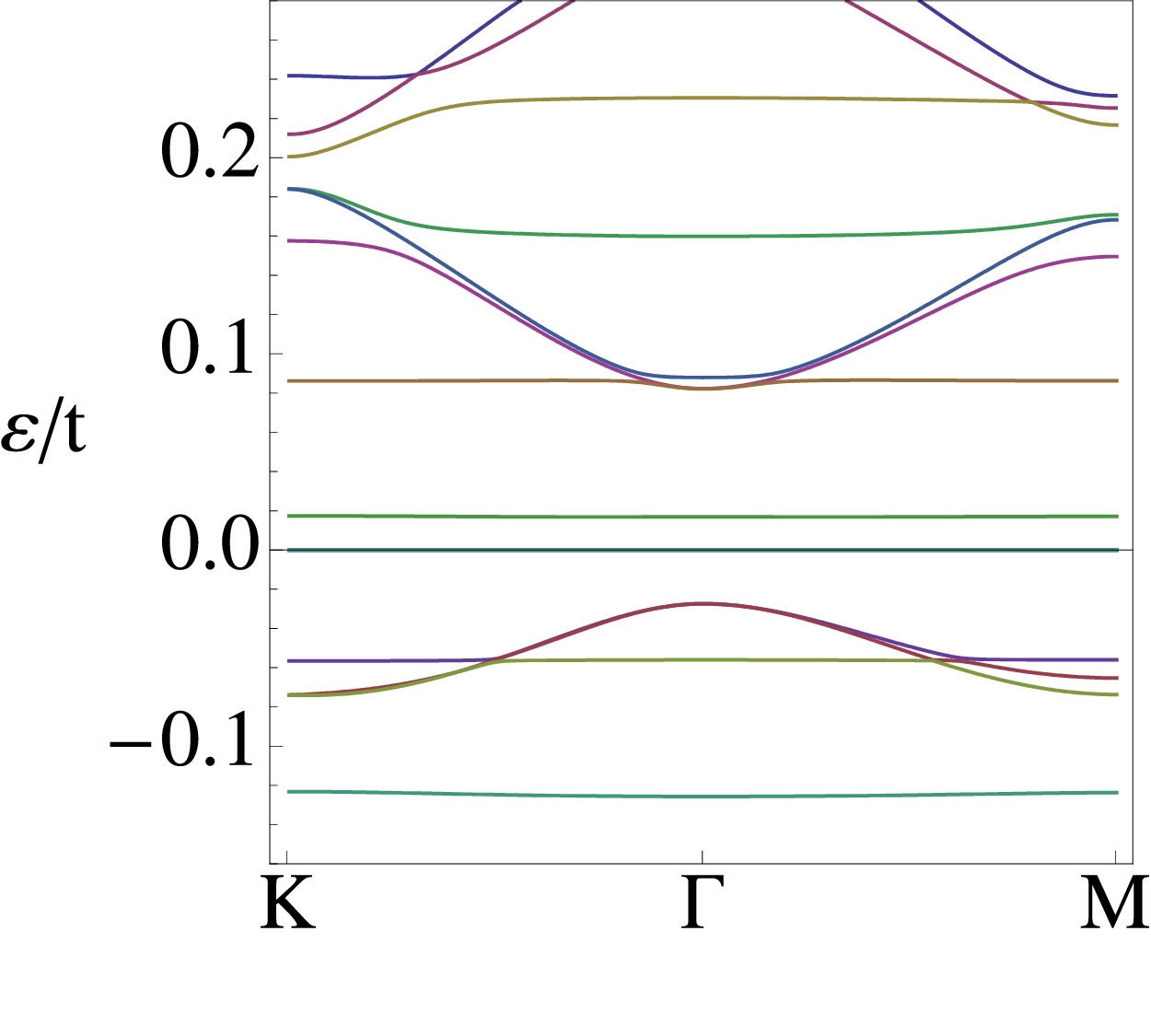}   \hspace{0.5cm}
\epsfxsize 5cm    \epsfbox{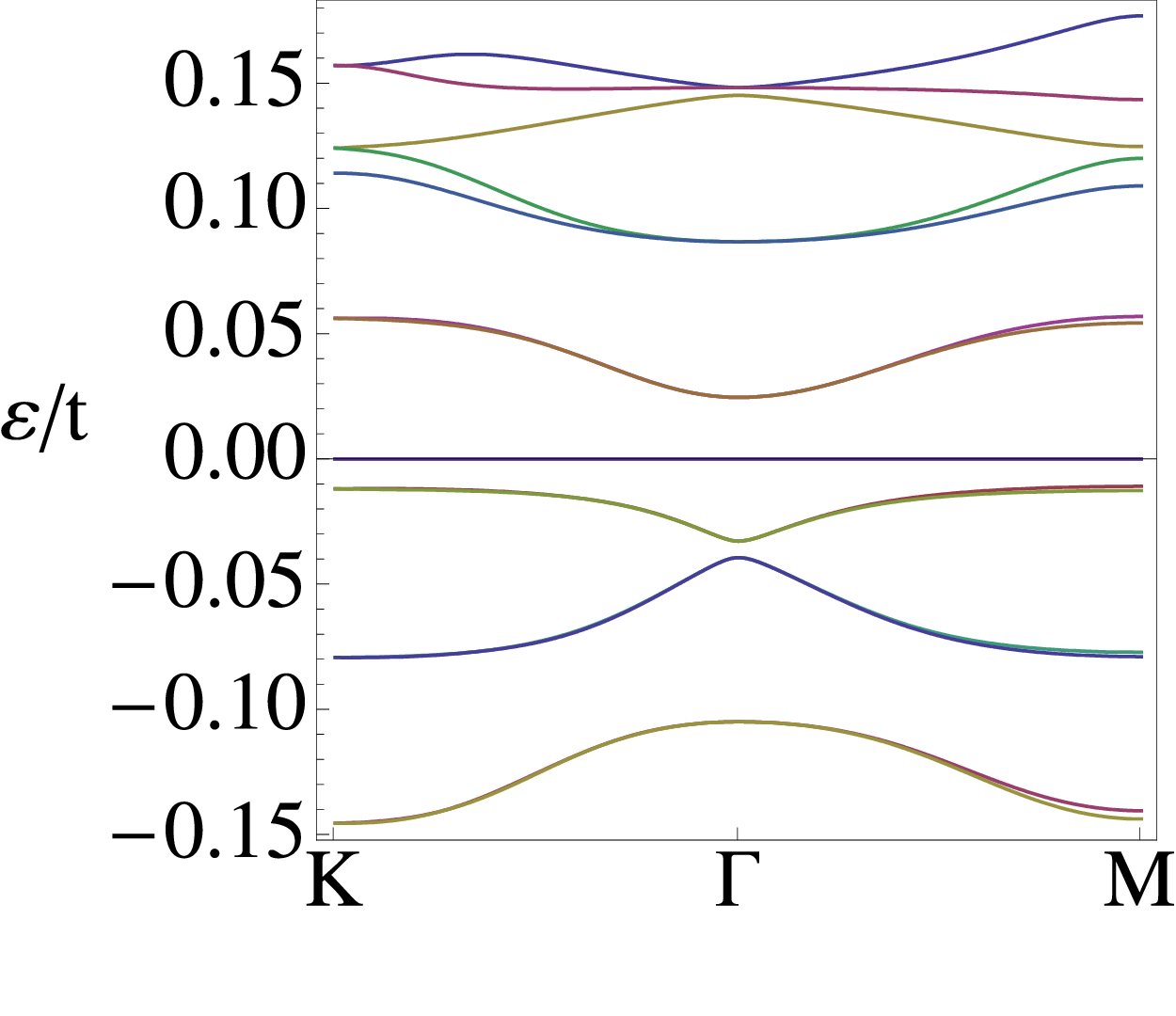} }\\
 \hspace{0.65cm}  (c) \hspace{4.75cm} (d)
\end{center}
\caption{Low-energy bands of arrays of junctions with nanotube geometries 
$(12,12)$ (a), $(12,0)$ (b), $(18,0)$ (c), and $(24,0)$ (d). In all the cases 
the unit cell of the array is of the type shown in Fig. \ref{one}, with a side 
of the hexagon in the basal plane equivalent to 10 carbon rings of the graphene 
sheet, and a nanotube height equivalent to 10 unit cells of the nanotube
for (a), (b), (c), and 20 nanotube unit cells for (d).}
\label{eight}
\end{figure}

The band structure of the array of junctions depends on the geometry of 
the nanotubes, as well as on their length and the distance between them. For
simplicity, we are going to consider arrays where all the nanotubes have the
same chirality. Then, it can be checked that the arrays fall into two main 
classes, regarding the behavior of the bands close to the Fermi
level. The distinctive feature of one class with respect to the other is the
presence of flat bands in the low-energy part of the spectrum. The arrays of
junctions made of armchair nanotubes, for instance, always have a number of 
these flat bands, as illustrated by the representative in Fig. \ref{eight}(a). 
The appearance of flat bands in a particular array of junctions was
noticed in Ref. \onlinecite{jap}. We have found that the flat bands are
actually generic in arrays made of armchair nanotubes, which display a series 
of them as one moves from the Fermi level to higher (or lower) energies. The 
spacing in energy between the flat bands becomes inversely proportional to 
the length of the nanotubes. The bands dispersing at low energies in Fig. 
\ref{eight}(a) are not affected however by variations of that variable, while 
they move instead closer to the Fermi level as the distance between the 
nanotubes in the array is increased.

On the other hand, the presence of flat bands at low energies is not generic
in arrays made of zig-zag nanotubes. In general, we may expect a number of 
bands dispersing above and below the Fermi level, as shown in Figs. 
\ref{eight}(b) and (d), which represent the low-energy bands of arrays made
respectively of $(12,0)$ and $(24,0)$ nanotubes. In the case of the zig-zag 
nanotubes, flat bands only appear close to the Fermi level when the junctions
are formed with $(6n,0)$ geometries such that $n$ is a multiple of 3. This 
distinctive behavior can be appreciated in Fig. \ref{eight}(c), which displays
the low-energy bands in the case of an array made of $(18,0)$ nanotubes. The 
shape of the bands resembles there the typical appearance of the spectra of 
arrays made of armchair nanotubes, as shown in Fig. \ref{eight}(a). 

The mentioned flat bands have their origin in the existence of localized 
states in the arrays of junctions. We have checked that the junctions made
of armchair nanotubes and $(6n,0)$ nanotubes with $n$ equal to a multiple of 3
have in common the formation of electron states confined mostly in the nanotube 
side. These are the states responsible for the development of the flat bands 
shown in Fig. \ref{eight}, as the wave functions with most of their weight in 
the nanotubes show little overlap in the graphene part of the lattice. This
also explains in a natural way the proliferation of flat bands at low energies
as the nanotube length is increased, by thinking of the confined modes as 
standing waves in the tube. On the other hand, the states that are 
preferentially localized at the junctions (corresponding to the quasi-bound 
states of the Dirac equation) may be identified here as the pairs
of branches degenerated at the $\Gamma $ point. These states may have in 
general a significant overlap between nearest junctions, which is reflected 
in the appreciable dispersion of the corresponding bands.

We can reach in the continuum limit a qualitative understanding of the similar
behavior of the arrays made of armchair and $(6n,0)$ nanotubes when $n$ is 
a multiple of 3, by noticing that these are the only geometries that support 
low-energy standing waves between the junction and the other end of the tube. 
This requires the superposition of two modes with opposite momenta along the
tube, which is possible at low energies in the armchair nanotubes as the modes
at opposite Dirac points have then vanishing angular momentum. In general,
this is not the case for the $(6n,0)$ geometries, since in the zig-zag 
nanotubes the Dirac points correspond to large momenta in the transverse 
direction. Yet the formation of standing waves is possible when $n$ is a 
multiple of 3, as the low-energy states about the two Dirac points fall then 
in the same sector with quantum number $q = 1$ regarding the $C_{6v}$ symmetry. 
Thus, it is possible to form a state confined in the nanotube by superposition 
of two modes with opposite longitudinal momenta and the same quantum number 
$q$. This is consistent with the fact that the confined states are actually 
found in the $q =1$ sector in the diagonalization of very large lattices of 
individual nanotube-graphene junctions. In the real lattice of the array, the 
confinement of the electrons in the nanotubes is only approximate, but the 
decay of the wave functions in the graphene part away from the junctions is 
strong enough to account for the development of the flat bands shown above.

\section{Conclusions}

In this paper we have studied the electronic structure of the hybrid material
made of carbon nanotubes attached to a graphene sheet. By analyzing individual
nanotube-graphene junctions, we have found the following features:

i) Low-energy electrons in the graphene layer, with 
$|\varepsilon | \lesssim v_F/R_0$, are scattered by the nanotube, and the 
probability of propagating into the tube is small.

ii) High-energy electrons reaching the nanotube junction, with 
$|\varepsilon | \gg v_F/R_0$, are mostly transmitted into the nanotube.

iii) At low energies, $|\varepsilon | \lesssim v_F/R_0$, and in the vicinity
of the junction, $r \sim R_0$, there is in general a depletion of the density 
of states.

iv) In certain nanotube geometries (armchair and $(6n,0)$ with $n$ equal to 
a multiple of 3), there are quasi-bound states near $\varepsilon = 0$, 
partially localized at the junction.

We have shown that these features can be accounted for in a continuum model
of the hybrid geometry. This is based on the Dirac fermion fields describing 
the electronic excitations, interacting with the curvature and the effective 
gauge field arising from the six heptagonal carbon rings at the junction. Thus, 
properties i), ii) and iii) are intrinsic to the continuum Dirac equation and
universal for all nanotube geometries, while iv) depends on the relative 
position of the six heptagonal rings and the consequent effective magnetic 
flux at the junction. While we have focused on the case of armchair and 
zig-zag nanotubes, it becomes clear that the continuum theory may account as 
well for the properties of junctions with other geometries. In this respect, 
it is likely that, by allowing for less regular distributions of the heptagonal 
rings, nanotubes with nontrivial helicity can also be attached to the graphene 
sheet. In a more general theoretical perspective, it may be interesting to 
analyze other discrete realizations of the 2D Dirac equation, like the 
geometry of a square lattice with one half magnetic flux per plaquette.

We have also shown that the arrays of nanotube-graphene junctions fall into
two main classes, depending on whether their spectra exhibit or not flat 
bands close to the Fermi level. The flat bands only appear in arrays made
of armchair nanotubes or $(6n,0)$ nanotubes when $n$ is a multiple of 3.
On the other hand, the semiconducting behavior seems to be a constant in 
the class characterized by the presence of the flat bands, as no dispersive 
bands cross then the Fermi level. Metallic behavior of the array of junctions 
is possible in the other class, as shown in Fig. \ref{eight}(b), though that
behavior does not appear to be a generic trend, as illustrated by the absence
of low-energy bands crossing the Fermi level in the other representative 
of the class shown in Fig. \ref{eight}(d).

In real experimental samples, it is quite likely that the arrays may be
formed by junctions with nanotubes of different helicities. In this case, 
we can expect that the electronic structure of these arrays will be a mixture
of the features already present in Figs. \ref{eight}(a)-(d). In particular,
part of the electronic states will be still confined in some of the nanotubes,
and other states will be partially localized at some of the junctions. The 
feasibility of using the arrays of nanotube-graphene junctions may depend on
the possibility to tailor these hybrid structures to get specific 
functions. At this point, more input from experimental measurements on these
arrays would be required, while the remarkable behavior predicted for
these systems (localization and confinement of states, flat bands) opens good 
perspectives in the investigation of novel electronic devices.

\section{Acknowledgements}
We acknowledge many helpful discussions with A. H. Castro Neto, who
also mentioned to us the existence of Ref.~\cite{F08}. We
acknowledge financial support from MEC (Spain) through grants
FIS2005-05478-C02-01, FIS2005-05478-C02-02 and CONSOLIDER CSD2007-00010, 
by the Comunidad de Madrid, through CITECNOMIK, CM2006-S-0505-ESP-0337.

\end{document}